\documentclass[a4paper,12pt]{article}
\usepackage[dvips]{graphics}
\usepackage{amssymb}
\usepackage{amsmath}
\usepackage{color}
\usepackage{pstricks}
\usepackage{graphicx}
\usepackage{dcolumn} 
\usepackage{multirow, booktabs}

\usepackage{cite}
\newcommand{\GeV}{\; \mathrm{GeV}}

\newcommand{\beq}{\begin{equation}}
\newcommand{\eeq}{\end{equation}}
\newcommand{\bea}{\begin{eqnarray}}
\newcommand{\eea}{\end{eqnarray}}
\newcommand{\lsp}{\tilde\chi^0_1}
\newcommand{\mlsp}{m_{\tilde\chi^0_1}}

\newcommand{\tanb}{\tan\beta}

\newcommand{\grav}{\widetilde{G}  }

\newcommand\iso[2]{\mbox{${}^{#2}${\rm #1}}}
\def\he#1{\iso{He}{#1}}

\def\li#1{\iso{Li}{#1}}
\def\b1#1{\iso{B}{1#1}}

\newcommand*{\myalign}[2]{\multicolumn{1}{#1}{#2}}

\makeatletter

\def\slash{\@ifnextchar[{\fmsl@sh}{\fmsl@sh[0mu]}}
\def\fmsl@sh[#1]#2{%
  \mathchoice
    {\@fmsl@sh\displaystyle{#1}{#2}}%
    {\@fmsl@sh\textstyle{#1}{#2}}%
    {\@fmsl@sh\scriptstyle{#1}{#2}}%
    {\@fmsl@sh\scriptscriptstyle{#1}{#2}}}
\def\@fmsl@sh#1#2#3{\m@th\ooalign{$\hfil#1\mkern#2/\hfil$\crcr$#1#3$}}
\makeatother

\newcommand{\twovector}[2]{\left(\begin{array}{c} #1\\ #2 \end{array}\right)}

\newcommand{\fourvector}[4]{\left(\begin{array}{c} #1\\ #2 \\ #3 \\ #4 \end{array}\right)}
\newcommand{\eightvector}[8]{\left(\begin{array}{c} #1\\ #2 \\ #3 \\ #4 \\ #5\\ #6 \\ #7 \\ #8\end{array}\right)}
\newcommand{\tworow}[2]{#1 & #2}
\newcommand{\twotwomatrix}[2]{\left(\begin{array}{cc} \tworow#1 \\ \tworow#2 \end{array}\right)}

\def\a{\alpha}
\def\b{\beta}
\def\c{\chi}
\def\d{\delta}
\def\e{\epsilon}

\def\g{\gamma}

\def\l{\lambda}
\def\m{\mu}
\def\n{\nu}

\def\r{\rho}
\def\s{\sigma}

\newcommand{\mgrav}{m_{\widetilde{G}}}

\newcommand{\mplanck}{\ensuremath{M_{\text{P}}}}

\newcommand{\suthree}{\ensuremath{\text{SU}(3)_{\text{c}}}}
\newcommand{\sutwo}{\ensuremath{\text{SU}(2)_{\text{L}}}}

\newcommand{\cf}[2]{\c_{#1}^{#2}}   
\newcommand{\cfb}[2]{\overline{\c}_{#1}^{#2}}   
\newcommand{\gen}[3]{T_{#1,\, #2 #3\,}}  

\newcommand{\gbalpha}[2]{{A}^{(#1)\,#2}}   

\def\slash#1{\rlap{\hbox{$\mskip 1 mu /$}}#1}   

\definecolor{darkgreen}{rgb}{0,.5,0}

\newcommand{\comarray}[2]{{\tt
\vspace*{-1mm}\\
\begin{tabular}{l@{\hspace*{2cm}}l}  
{#1} & (Fortran) \\
{#2} & (Mathematica)
\end{tabular}\\
\vspace*{2mm}\\}}

\begin{document} 
\begin{titlepage} 
 
\begin{flushright}
HEPHY-PUB 959/15\\  
 \end{flushright} 
\begin{center} 
\vspace*{1.5cm} 
 
{\Large{\textbf {  {\tt \bf GravitinoPack} and decays of supersymmetric metastable particles}}}\\ 
 \vspace*{10mm} 
 
{\bf  Helmut~Eberl}$^1$ and  {\bf  Vassilis~C.~Spanos}$^2$ \\ 

\vspace{.7cm} 

$^1${{\it Institut f\"ur Hochenergiephysik der \"Osterreichischen Akademie
der Wissenschaften,}\\
{\it A--1050 Vienna, Austria}

 $^2${\it University of Athens, Faculty of Physics, Department of Nuclear {\rm \&} Particle Physics, \\ GR--15784 Athens, Greece}} \\

\end{center} 
\vspace{2.cm}
 
\begin{abstract} 
We present the package {\tt {GravitinoPack}} that calculates 
the two- and three-body decays of unstable supersymmetric particles involving the gravitino in
the final or initial state. 
In a previous paper, we already showed results for the gravitino decays into two and three particles.
In this paper, we incorporate the processes where an unstable neutralino, stau or stop decays into a gravitino
and Standard Model particles. This is the case in gravitino dark matter supersymmetric models, where  
the gravitino is the lightest SUSY particle.
We give instructions for the installation and the use of the package. In the numerical analysis, we discuss various MSSM scenarios.
We show that the calculation of all the decay channels and the three-body decay branching ratios is essential for 
the accurate application of cosmological bounds on these models.
\end{abstract}

\end{titlepage} 

\baselineskip=18pt
\section{Introduction}
\label{intro}

In the past the gravitino dark matter (DM) scenario has been studied extensively~\cite{decays_old}. 
In these scenarios assuming $R$-parity conservation,
the gravitino, the supersymmetric partner of the graviton, is stable and can play the role of the DM particle. 
Other supersymmetric particles as neutralinos and sfermions (e.g. stops or staus) are unstable and can decay into a
gravitino and other Standard Model (SM) particles. 
These decays produce electromagnetic energy and hadrons which
affect the primordial Big-Bang
Nucleosynthesis (BBN) prediction for the  abundances of the light nuclei, 
like D, \he4, \he3 and \li7~\cite{decays_new , cefo_etc, cefos}.

In a previous paper~\cite{Gravitino_decays}  we presented results for the complementary case where the
gravitino is unstable, using the package {\tt GravitinoPack}. 
In this paper we extend the scope of the package by calculating the decays widths and the branching ratios 
of decays of unstable supersymmetric particles into a gravitino. 

The package {\tt GravitinoPack} is a numerical tool developed with the help of the  packages
{\tt  FeynArts} ({\tt FA}) and {\tt FormCalc} ({\tt FC}),~\cite{FAref,  FAFCref , LTref }. 
It contains {\sc Fortran}  and  {\sc Mathematica} routines that calculate the decay rates for the
main decay channels for the unstable supersymmetric particles, involving gravitinos.  
In~\cite{Gravitino_decays} we studied in detail all dominant two-body 
channels $\grav \to \widetilde{X}\, Y$, as
well as the three-body channels  $ \grav  \to  \lsp \, X \,Y $,   where $\widetilde{X}$ is a sparticle, $\lsp$ the lightest 
neutralino and $X$, $Y$ are SM particles. The two-body
decays dominate the total gravitino width, and in particular the channel $\grav \to  \lsp \, \gamma $,  which  is kinematically open 
in the whole region $\mgrav > \mlsp$. On the other hand, 
also many three-body decay channels can be open, $ \grav  \to  \lsp \, X \,Y $, even below thresholds of involved
two-body decays, $\mgrav < m_{\widetilde{X}}  + m_Y$.

In this paper we present results for the cases where 
the Next to the Lightest  Supersymmetric Particle (NLSP) is the lightest neutralino ($\lsp$), 
or the lighter stau ($\tilde \tau_1$), or the lighter stop ($\tilde t_1$) decaying to a gravitino and Standard Model particles.
The dominant two-body decay channels are $\lsp \to \grav \, \gamma \, (Z)$, $ \tilde \tau_1 \to  \grav \, \tau  $ and $ \tilde t_1 \to  \grav \, t $.
In addition, we have included all possible three-body decays.
 
The amplitudes of the processes are generated using the package  {\tt FA}, that has been  extended 
in order to deal with interactions with spin-3/2 particles. We have built a model file with all possible gravitino interactions 
with the particles of the Minimal Supersymmetric Standard Model (MSSM). The corresponding details are shown in Appendix~A.
{\tt FC} has been extended so that it automatically generates a {\sc Fortran} code for the numerical calculation
of the squared amplitudes. As there are many gamma matrices involved we have done this within {\tt FC} by using the Weyl-van-der-Waerden 
formalism~\cite{WvdW} as implemented into {\tt FC} from~\cite{Dittmaier:1998nn}. There the complexity of the calculation only grows 
linearly with the number of Feynman graphs.
 
In our numerical study, we use a few benchmark points from supersymmetric models with different supersymmetry breaking patterns,
like the phenomenological MSSM (pMSSM)~\cite{pMSSM},   and the 
Constrained MSSM (CMSSM)~\cite{cmssm1,cmssm2}. Actually, the cases we have selected 
in the pMSSM are mainly points where the neutralino carries significant Higgsino components as in the  
Non-Universal Higgs Model (NUHM)~\cite{nuhm}. 
Therefore, we do not explicitly discuss the NUHM. These models are applied 
to the different NLSP cases. 
In our analysis, we have employed all phenomenological constraints  from the LHC~\cite{LHC} experiments 
concerning  the superpartner  mass bounds,
the Higgs boson mass~\cite{mh125} and  the LHCb~\cite{bmm,LHCb} data.  

The paper is organized as follows: In Section~2 we present the decay channels into the LSP gravitino.
In Section~3 we describe the package {\tt GravitinoPack} and give details for its  installation and use. 
We present in Section~4 a few representative numerical results in various MSSM models. 
In Section~5 we summarise our results.
In Appendix~\ref{Gr_interactions} the detailed derivation of the gravitino couplings to the MSSM particles is given.


\section{The Calculation}

The main aspects were already presented in \cite{Gravitino_decays}.
There an illustrative example is given how a specific gravitino-MSSM interaction is derived and
implemented into {\tt FeynArts}
 together with all possible Lorentz structures of gravitino
interactions with MSSM particles. The detailed derivation of the total explicit gravitino-MSSM 
interaction Lagrangian with the 78 couplings can be found in Appendix~\ref{Gr_interactions}.  

\subsection{Two- and Three-body decays with Gravitino}

In \cite{Gravitino_decays} we already discussed all two-body decays of the gravitino and
all three-body of gravitino into a neutralino and a SM particle pair.
We also present the general formulas for the decay widths.
The analogous formulas for the decays into a gravitino can be derived from there.

\subsubsection{NLSP neutralino decays}

There are five two-body decays of $\tilde \chi_1^0$ possible:
\begin{eqnarray*}
\widetilde \chi_1^0  & \to  & \tilde G \,(Z^0\, ,\gamma) \, ,\nonumber \\
\widetilde \chi_1^0  & \to  & \tilde G \,(h^0, H^0, A^0)  \, .
\label{neu2Gr two-body decays}
\end{eqnarray*}
\noindent
The lightest neutralino decays into the gravitino and a pair of SM particles as
\begin{eqnarray*}
\widetilde \chi_1^0  & \to  & \tilde G \, \bar f f \,, \nonumber \\
\widetilde \chi_1^0  & \to  & \tilde G \, V V\, ,\quad V V = Z^0 Z^0\,, Z^0 \gamma\,, W^+ W^- \, ,\nonumber \\
\widetilde \chi_1^0  & \to  & \tilde G \, V S\, ,\quad V S = (Z^0, \gamma)(h^0, H^0, A^0), W^+ H^-, W^- H^+\,, \nonumber \\
\widetilde \chi_1^0  & \to  & \tilde G \, S S\, ,\quad S S = (h^0, H^0, A^0) (h^0, H^0, A^0), H^+ H^-\,,
\label{neu2Gr three-body decays}
\end{eqnarray*}
where $f  =  \nu_e, \,\nu_\mu,\, \nu_\tau,\,  e^-,\, \mu^-, \, \tau^-, \,  u,\, c,\, t,\, d,\, s,\, b$.
These are 19 three-body decay channels. They are given in Table~\ref{tab:neu2Gr3bodydecays}.
\renewcommand{\arraystretch}{1.3}
\begin{table}[h!]
\begin{center}
$\begin{array}{|l|c|l|l|}
\hline
\myalign{|c}{\rm process} & \myalign{|c}{\rm number} & \myalign{|c}{\rm first~decay} & \myalign{|c|}{\rm possible} \\[-2mm]
\myalign{|c}{\widetilde \chi^0_1 \to \tilde G X Y} & \myalign{|c}{\rm of~graphs} & \myalign{|c}{\widetilde \chi^0_1 \to \tilde X Y} & \myalign{|c|}{\rm resonances}\\
\hline 
\hline
\tilde G f \bar f  & 7 & \tilde G (h^0 , H^0 , A^0 , \gamma , Z^0 ) , f \tilde f^*_l , \tilde f_l \bar f  & h^0 , H^0 , A^0 , Z^0 \\ 
\tilde G Z^0 Z^0  & 4 & \tilde G (h^0 , H^0 ) , Z^0 \tilde \chi_k^0 , \tilde \chi_k^0 Z^0  & H^0 \\ 
\tilde G Z^0 \gamma  & 1 & \tilde \chi_k^0 Z^0 & - \\ 
\tilde G W^+ W^-  & 6 + 4pt  & \tilde G (h^0 , H^0 , \gamma , Z^0 ) , W^+ \tilde \chi_j^- , \tilde \chi_j^+ W^-  & H^0\\ 
\tilde G Z^0 h^0  & 4 + 4pt  & \tilde G (A^0 , Z^0 ) , Z^0 \tilde \chi_k^0 , \tilde \chi_k^0 h^0  & A^0  \\ 
\tilde G Z^0 H^0  & 4 + 4pt  & \tilde G (A^0 , Z^0 ) , Z^0 \tilde \chi_k^0 , \tilde \chi_k^0 H^0  & A^0  \\ 
\tilde G Z^0 A^0  & 4 + 4pt  & \tilde G (h^0 , H^0 ) , \tilde \chi_k^0 Z^0 , A^0 \tilde \chi_k^0  & H^0   \\ 
\tilde G \gamma h^0  & 1 & \tilde \chi_k^0 h^0  & - \\ 
\tilde G \gamma H^0  & 1 & \tilde \chi_k^0 H^0 & - \\ 
\tilde G \gamma A^0  & 1 & \tilde \chi_k^0 A^0  & - \\ 
\tilde G W^+ H^-  & 5 + 4pt  & \tilde G (h^0 , H^0 , A^0 ) , W^+ \tilde \chi_j^- , \tilde \chi_j^+ H^-  & H^0 , A^0  \\ 
\tilde G W^- H^+  & 5 + 4pt  & \tilde G (h^0 , H^0 , A^0 ) , W^- \tilde \chi_j^+ , \tilde \chi_j^- H^+  & H^0 , A^0  \\ 
\tilde G h^0 h^0  & 4 & \tilde G (h^0 , H^0 ) , h^0 \tilde \chi_k^0 , \tilde \chi_k^0 h^0  & H^0   \\ 
\tilde G H^0 H^0  & 4 & \tilde G (h^0 , H^0 ) , H^0 \tilde \chi_k^0 , \tilde \chi_k^0 H^0  & - \\ 
\tilde G h^0 H^0  & 4 & \tilde G (h^0 , H^0 ) , h^0 \tilde \chi_k^0 , \tilde \chi_k^0 H^0  &  - \\ 
\tilde G A^0 A^0  & 4 & \tilde G (h^0 , H^0 ) , A^0 \tilde \chi_k^0 , \tilde \chi_k^0 A^0  & H^0  \\ 
\tilde G h^0 A^0  & 3 & \tilde G (A^0 , Z^0 ) , h^0 \tilde \chi_k^0   & - \\ 
\tilde G H^0 A^0  & 3 & \tilde G (A^0 , Z^0 ) , H^0 \tilde \chi_k^0  & - \\ 
\tilde G H^+ H^-  & 6 & \tilde G (h^0 , H^0 , \gamma , Z^0 ) , H^+ \tilde \chi_j^- , \tilde \chi_j^+ H^-  & H^0  \\ 
\hline  
\end{array}$
\end{center}
\renewcommand{\arraystretch}{1}
\caption[3body decay table]{All possible three-body decays 
channels of the NLSP neutralino $\widetilde \chi^0_1$ decaying into the LSP gravitino $\tilde G$ and a pair of SM particles; $4pt$ denotes a Feynman graph
with four-point interaction. The indices are $i = 1, \ldots, 4$; $k = 2,3,4$; $j,l = 1,2$, and
$f  =  \nu_e, \,\nu_\mu,\, \nu_\tau,\,  e^-,\, \mu^-, \, \tau^-, \,  u,\, c,\, t,\, d,\, s,\, b$.}
\label{tab:neu2Gr3bodydecays}
\end{table}

\subsubsection{NLSP stau decays}

There is only one two-body decay possible,
\begin{eqnarray*}
\widetilde \tau_1^- & \to  & \tilde G \,\tau^-  \, .\\
\label{stau2Gr two-body decays}
\end{eqnarray*}
\noindent
All possible three-body decays of $\widetilde \tau_1^-$ are given in Table~\ref{tab:stau2Gr3bodydecays}.
\renewcommand{\arraystretch}{1.3}
\begin{table}[h!]
\begin{center}
$\begin{array}{|l|c|l|}
\hline
\myalign{|c}{\rm process} & \myalign{|c}{\rm number} & \myalign{|c|}{\rm first~decay} \\[-2mm]
\myalign{|c}{\tilde \tau_1 \to \tilde G X Y} & \myalign{|c}{\rm of~graphs} & \myalign{|c|}{\tilde \tau_1 \to \tilde X Y}\\
\hline 
\hline
\tilde G Z^0 \tau & 3 + 4pt & \tilde G \tau,\, \tilde \tau_i Z^0,\, \tilde \chi_k^0 \tau\\
\tilde G W^- \nu_\tau & 3 + 4pt & \tilde G \tau,\, \tilde \nu_\tau W^-,\, \tilde \chi_j^- \nu_\tau\\
\tilde G h^0 \tau & 3 & \tilde G \tau,\, \tilde \tau_i h^0 ,\, \tilde \chi_k^0 \tau\\
\tilde G H^0 \tau & 3 & \tilde G \tau,\, \tilde \tau_i H^0,\, \tilde \chi_k^0 \tau\\
\tilde G A^0 \tau & 3 & \tilde G \tau,\, \tilde \tau_i A^0,\, \tilde \chi_k^0 \tau\\
\tilde G H^- \nu_\tau & 3 & \tilde G \tau,\, \tilde \nu_\tau H^-,\, \tilde \chi_j^- \nu_\tau\\
\hline  
\end{array}$
\end{center}
\renewcommand{\arraystretch}{1}
\caption[3body decay table]{All possible three-body decays 
channels of the NLSP stau $\tilde \tau_1$ decaying into the LSP gravitino $\tilde G$ and a pair of SM particles; $4pt$ denotes a Feynman graph
with four-point interaction. The indices are $i, j = 1,2$ and $k = 1,2,3,4$. There are no resonances possible.}
\label{tab:stau2Gr3bodydecays}
\end{table}

\subsubsection{NLSP stop decays}

There is only one two-body decay possible,
\begin{eqnarray*}
\widetilde t_1 & \to  & \tilde G \,t  \, .\\
\label{st2Gr2body decays}
\end{eqnarray*}
\noindent
All possible three-body decays of $\tilde t_1$ are given in Table~\ref{tab:stop2Gr3bodydecays}. If 
$\tilde t_1$ is the NLSP only the top quark can be resonant in the channels with $W^+$ or $H^+$. 
\renewcommand{\arraystretch}{1.3}
\begin{table}[h!]
\begin{center}
$\begin{array}{|l|c|l|l|}
\hline
\myalign{|c}{\rm process} & \myalign{|c}{\rm number} & \myalign{|c|}{\rm first~decay}& \myalign{|c|}{\rm possible} \\[-2mm]
\myalign{|c}{\tilde t_1 \to \tilde G X Y} & \myalign{|c}{\rm of~graphs} & \myalign{|c|}{\tilde t_1 \to \tilde X Y} & \myalign{|c|}{\rm resonances}\\
\hline 
\hline
\tilde G Z^0 t & 3 + 4pt & \tilde G t,\, \tilde t_i Z^0,\, \tilde \chi_k^0 t & \tilde \chi_k^0 \\
\tilde G W^+ b & 3 + 4pt & \tilde G t,\, \tilde b_i  W^+,\, \tilde \chi_j^+ b & t, \tilde b_i, \tilde \chi_j^+ \\
\tilde G h^0 t & 3 & \tilde G t,\, \tilde t_i h^0 ,\, \tilde \chi_k^0 t & \tilde \chi_k^0 \\
\tilde G H^0 t & 3 & \tilde G t,\, \tilde t_i H^0,\, \tilde \chi_k^0 t & \tilde \chi_k^0 \\
\tilde G A^0 t & 3 & \tilde G t,\, \tilde t_i A^0,\, \tilde \chi_k^0 t & \tilde \chi_k^0\\
\tilde G H^+ b & 3 & \tilde G t,\, \tilde b_i H^+,\, \tilde \chi_j^+ b & t, \tilde b_i, \tilde \chi_j^+ \\
\hline  
\end{array}$
\end{center}
\renewcommand{\arraystretch}{1}
\caption[3body decay table]{All possible three-body decays 
channels of the stop $\tilde t_1$ decaying into the LSP gravitino $\tilde G$ and a pair of SM particles; $4pt$ denotes a Feynman graph
with four-point interaction. The indices are $i, j = 1,2$ and $k = 1,2,3,4$.}
\label{tab:stop2Gr3bodydecays}
\end{table}

%
%

\section{\textbf{{\tt GravitinoPack}}}
{\tt GravitinoPack} is a package for the evaluation of processes with gravitino interaction.
The version {\tt GravitinoPack1.0} includes all two-body decays of $\tilde G$ and all 
three-body decays of $\tilde G$ to a neutralino and a pair of two particles. In the
case the gravitino is the LSP all two- and three-body decays of the $\tilde \chi^0_1$, $\tilde \tau_1$ or $\tilde t_1$  NLSP
are included. {\tt GravitinoPack} works at the Fortran level and has a Mathematica interface.
All two-body decay widths with a gravitino were cross-checked with the analytic results of 
the package {\tt FeynRules}~\cite{FeynRules2}, version 2.0.23. We have found agreement in all channels. 
In {\tt GravitinoPack} the input can also be given in SUSY Les Houches Accord convention \cite{SLHAref,SLHA2ref}. 
For convenience we have directly included the code of the libraries SLHALib-2.2~\cite{slhalib},
and Cuba-3.3~\cite{Cubalib} for the three-body phase space integrations.\\

\subsection{Installation}

To compile the package, a Fortran 77 compiler and the GNU C compiler (gcc) are required.

\begin{enumerate}
\item Download the file {\tt GravitinoPack1.0.tar.gz} at
\begin{quote} http://www.hephy.at/susytools \end{quote}
\item Unpack the archive by
\begin{quote} {\tt ar -xvzf GravitinoPack1.0.tar.gz} \end{quote}
\item Go to the {\tt GravitinoPack1.0} folder and write
\begin{quote} {\tt ./configure} \end{quote}
which creates the {\tt makefile}
 
\item Fortran programs with the main fortran file {\tt example1.F} can be compiled by 
\begin{quote} {\tt make example1} \end{quote}
\item That will generate an executable called {\tt example1}. To run it type
\begin{quote} {\tt ./example1} \end{quote}
\item The Mathematica link program {Mgravitinopack} is compiled by 
\begin{quote} {\tt make Mgravitinopack} \end{quote}
\end{enumerate}

\subsection{Use of {\tt GravitinoPack}}

In the main directory of the package there are five examples in order to explain 
the functionality at the Fortran level, see the main files {\tt example1.F}, {\tt example1slha.F},
{\tt example2slha.F},  {\tt example3slha.F}, and  {\tt example4slha.F}.
Furthermore, the file {\tt GravitinoPack.nb} shows the usage at the Mathematica level,
working with the MathLink executable {\tt Mgravitinopack}. Executing the command {\tt make} 
creates all six executables at once.\\

\noindent
{\tt GravitinoPack} works at tree-level with possible complex
flavor conserving parameters. The input can be set locally or read in from a SLHA file. 
Then decay widths and branching ratios can be calculated.\\ 

\noindent
First of all, one has to define a scenario. We set the SM parameters given 
in {\tt setSMparameters.F}:\\
\comarray{call setSMparameters()}{SetSMparameters[]}
If a SLHA file is read in, this call becomes redundant.\\

\noindent
Then the input parameters vector with 43 entries must be set, see also the code in {\tt example1.F}, \\
{\tt input = \{MA0, TB, absM1, phiM1, M2, absMUE, phiMUE, MSQ[1], MSQ[2], MSQ[3], 
 MSU[1], MSU[2], MSU[3], MSD[1], MSD[2], MSD[3], MSL[1], MSL[2], 
 MSL[3], MSE[1], MSE[2], MSE[3], absAe[1], phiAe[1], absAe[2], 
 phiAe[2], absAe[3], phiAe[3], absAu[1], phiAu[1], absAu[2], phiAu[2],
  absAu[3], phiAu[3], absAd[1], phiAd[1], absAd[2], phiAd[2], 
 absAd[3], phiAd[3], absM3, phiM3, MGr\}}\\ 

\noindent
Using a SLHA file, one reads in the input parameters vector by\\
\comarray{call getslhapara(slhainfile, input)}{input = GetSLHAParameters[slhainfile]}
The gravitino mass $m_{\tilde G}$ must be set in addition,\\
\comarray{input(43) =   $m_{\tilde G}$}{input[[43]] = $m_{\tilde G}$}

\noindent
All parameters including SUSY masses, mixing angles and total widths of possible resonant propagators in three-body
decays are calculated and some default values are set by\\
\comarray{call setMSSMparameters(input, flag)}{SetMSSMparameters[input, flag]}
with 
{\tt flag = \{flag1, flag2\}} where\\
{\tt flag1}  = 1/0: print information on/off,\\
{\tt flag2}   = 1/0 : gauge unification on/off.\\

\noindent
Working with a SLHA input file, {\tt flag2} becomes redundant, and one further can use\\
\noindent
\comarray{call setMSSMOSparameters()}{SetMSSMOSparameters[]}
which reads from the {\tt slhainfile} file all on-shell SUSY and Higgs masses, the 
rotation matrices of charginos, neutralinos, stops, sbottoms and staus, 
the $h^0 - H^0$ mixing angle $\alpha$, and
all the widths for possibly resonant propagators.\\
\comarray{call setMSSMOSmasses()}{SetMSSMOSmasses[]}
takes from the SLHA file all on-shell SUSY and Higgs masses.\\

\noindent
The Mathematica variable 
{\tt SLHAfile} shows the currently read LesHouches file. 
Further useful functions are 
{\tt GetSMparameters[]} and {\tt GetMSSMmasses[]}, see the description for them directly in 
{\tt examples.nb}. \\

\noindent
Now we come to the functions which calculate decay widths given in GeV and branching ratios (BRs).
Note, in the following {\tt A} denotes the photon and  {\tt HH} stands for the heavy CP even Higgs $H^0$.
All the other particle names in the code should be self-explanatory.\\

\noindent
For the decays there are five couples of miscellaneous functions given, which all work basically on the
same principle. By calling {\tt SetMSSMparameters} the default values of the corresponding flags  are set, see Table~{\ref{table:set3bdefaults}}.

\begin{table}[h!]
\begin{center}
\begin{tabular}{|c|c|}
\hline
flag & default value\\
\hline
\hline
{\tt itwobody}  & 1\\
\hline
{\tt iCMS}  & 1\\
{\tt ithreebody} & 1\\
 {\tt ionlygamma}  & 0\\
 \hline
{\tt igrhel} & 0\\
\hline
\end{tabular}
\caption{{\tt SetMSSMparameters} sets the default values of five flags.  
\label{table:set3bdefaults}}
\end{center}
\end{table}

\noindent
\comarray{call seti2body[i]}{Seti2body[i]} 
sets the flag {\tt itwobody = i}.\\[5mm]
{\tt itwobody = 0}: analytic formulas are used,\\
{\tt itwobody = 1}: FormCalc results are used.\\ 
\comarray{getitwobody()}{Geti2body[]} returns the actual value of {\tt itwobody}.

\noindent
\comarray{call setiCMS(i)}{SetiCMS[i]}
sets the flag {\tt iCMS = i} for the frame used for the three-body center-of-mass system,
{\tt iCMS}  can be 1,2, or 3.\\
\comarray{getiCMS()}{GetiCMS[]} returns the actual value of {\tt iCMS}.

\noindent
\comarray{call seti3body[i]}{Seti3body[i]} 
sets the flag {\tt ithreebody = i}.\\[5mm]
{\tt ithreebody = 0}: full calculation, also in case of resonances\\
{\tt ithreebody = 1}: NWA for resonances and non-resonant part\\ 
{\tt ithreebody = 2}: narrow width approx. for resonances only\\
{\tt ithreebody = 3}: non-resonant part only.\\ 
\comarray{getithreebody()}{Geti3body[]} returns the actual value of {\tt ithreebody}.\\
\comarray{call setionlygamma(i)}{Setionlygamma[i]} 
sets the flag {\tt ionlygamma = i}.\\
{\tt ionlygamma = 0}: all possible Feynman graphs are taken in decays with a $W^+ W^-$ pair.\\
{\tt ionlygamma = 1}: only the Feynman graph with photon line and four-point interaction are taken 
in decays with a $W^+ W^-$ pair. 

\noindent
\comarray{getionlygamma()}{Getionlygamma[]} 
returns the actual value of {\tt ionlygamma}. \\ 
\comarray{call setiGrHel[i]}{SetiGrHel[i]}
sets {\tt igrhel = i}. It affects only decays of ${\tilde G}$ and
works only for {\tt itwobody} = 1.\\
{\tt igrhel = 0}: unpolarized gravitino decay\\
{\tt igrhel = 1}: only spin 1/2 + spin -1/2 contributions are taken\\
{\tt igrhel = 3}: only spin 3/2 + spin -3/2 contributions are  taken\\
\comarray{getigrhel()}{GetiGrHel[]}
returns the actual value of {\tt igrhel}. \\

\noindent
Furthermore, the functions\\ 
\comarray{IsGrtheLSP()}{IsGrtheLSP[]}\\[-10mm]
\comarray{IsNeu1theNLSP()}{IsNeu1theNLSP[]}\\[-10mm]
\comarray{IsStau1theNLSP()}{IsStau1theNLSP[]}\\[-10mm]
\comarray{IsStop1theNLSP()}{IsStop1theNLSP[]}\\[-8mm]
are useful for the decays of  $\tilde \chi^0_1$, $\tilde \tau_1$, and  $\tilde t_1$ into ${\tilde G}$.
If the return value is 1, the condition is fulfilled, otherwise not. \\ 
\comarray{call gravwidth2body(args)}{GravWidth2body[] }
returns the branching ratios (BRs) and the total width of the gravitino decaying into all possible two-body
final states,
{\tt \{args\} = \{BR\_NeuA[4], BR\_NeuZ[4], BR\_GluinoG, BR\_ChaW[2], BR\_FSF[24], BR\_NeuHn[12], BR\_ChaH[2], gamma2tot\}},
{\tt A} denotes the photon.
The Fortran BR of the gravitino to fermion sfermion, {\tt BR\_FSF(i(2),type(4),gen(3))} is mapped into  
{\tt {BR\_FSF[1,1,1], BR\_FSF[2,1,1], BR\_FSF[1,2,1], BR\_FSF[2,2,1], $\ldots$, BR\_FSF[2,4,3]}}, 
{\tt type} = \{1,2,3,4\} == \{sneutrino, slepton, sup, sdown\} type.
Similar is the mapping of {\tt BR\_NeuH0: BR\_Neuh0} == elements \{1,2,3,4\},
{\tt BR\_NeuH0} == elements \{5,6,7,8\}, and  {\tt BR\_NeuA0} == elements \{9,10,11,12\} of {\tt BR\_NeuHn}.
Note that for the decays into a neutralino and a pair of charged particles the charged conjugated
channel is already included in the BR, e.g. {\tt BR\_ChaW} == {\tt BR\_ChapWm + BR\_ChamWp}.\\
\comarray{GravWidth3body(idecnumber, ineu, igen)}{GravWidth3body[idecnumber, ineu, igen]}
returns a certain gravitino three-body decay width to a neutralino and a pair of non-SUSY particles.
{\tt  ineu} = 1,2,3,4; for the decay to neutralino and a fermion pair also {\tt igen} = 1,2,3 must be set,
otherwise it is a dummy argument.
 {\tt idecnumber} stands for the gravitino decay to  
{\tt  \{\{NeuNNb, 1\}, \{NeuLLb, 2\}, \{NeuUUb, 3\}, \{NeuDDb, 4\}, 
 \{NeuWmWp, 5\}, \{NeuWpWm, 6\},\{NeuHpHm, 26\}, \{NeuHmHp, 27\}, \{NeuZZ, 7\}, 
 \{NeuAZ, 8\}, \{NeuZA, 9\}, \{Neuh0A, 10\}, \{NeuAh0, 11\}, \{NeuHHA, 12\}, \{NeuA0A, 13\},
 \{Neuh0Z, 14\}, \{NeuZh0, 15\}, \{NeuHHZ, 16\}, \{NeuA0Z, 17\}, \{NeuHpWm, 18\}, \{NeuWpHm, 19\},
\{Neuh0h0, 20\}, \{NeuHHHH, 21\}, \{NeuA0A0, 22\}, \{Neuh0HH, 23\}, \{Neuh0A0, 24\}, \{NeuHHA0, 25\}\}}.\\
\comarray{call gravtotalwidth(k, gam2tot, gam3tot)}{GravTotalWidth[k]}
calculates the total two-body decay width {\tt gam2tot} and the non-resonant three-body decay width into SM particle pairs and $\tilde \chi^0_k$, 
{\tt gam3tot}, with $k = 1,\ldots, 4$. For $k = 0$ all four contributions are summed up in the total non-resonant three-body decay width. 
The Mathematica result is given as {\tt \{gam2tot, gam3tot\}}.\\
 \comarray{call neu1toGrwidth2body(args)}{Neu1toGrWidth2body[] }
 returns all possible two-body BRs 
 and the total two-body width of $\tilde \chi^0_1$ decaying into $\tilde G$ and a SM particle, assuming that $\tilde \chi^0_1$  is the NLSP and Gr the LSP,
 {\tt \{args\} = \{BR\_GrA, BR\_GrZ, BR\_Grh0, BR\_GrH0,  BR\_GrA0, gamma2tot\}}.\\
\comarray{Neu1toGrWidth3body(idecnumber, igen)}{Neu1toGrWidth3body[idecnumber, igen]}
returns the width of a certain $\tilde \chi^0_1$ three-body decay to $\tilde G$ and a pair of non-SUSY particles, 
assuming that $\tilde \chi^0_1$  is the NLSP and Gr the LSP.
For the decay to $\tilde G$ and  a fermion pair {\tt igen} = 1,2,3 must be set,
otherwise it is a dummy argument.
{\tt idecnumber} stands for the $\tilde \chi^0_1$  decay to  
{\tt \{\{Neu2GrNNb, 1\}, \{Neu2GrLLb, 2\}, \{Neu2GrUUb, 3\}, \{Neu2GrDDb, 4\},
 \{Neu2GrWpWm, 6\}, \{Neu2GrZZ, 7\}, \{Neu2GrZA, 9\},
 \{Neu2GrZh0, 15\}, \{Neu2GrZHH, 16\}, \{Neu2GrZA0, 17\}, 
 \{Neu2GrAh0, 10\}, \{Neu2GrAHH, 12\}, \{Neu2GrAA0, 13\}, \{Neu2GrWpHm, 19\},
 \{Neu2Grh0h0, 20\}, \{Neu2GrHHHH, 21\}, \{Neu2GrA0A0, 22\}, \{Neu2Grh0HH, 23\}, \{Neu2Grh0A0, 24\}, \{Neu2GrHHA0, 25\},
 \{Neu2GrHpHm, 26\}\}}.
  
\noindent
\comarray{call neu1toGrtotalwidth(gam2tot, gam3tot)}{Neu1toGrTotalWidth[]}
calculates the total two-body decay width  {\tt gam2tot} and the non-resonant three-body decay width into SM particle pairs
of $\tilde \chi^0_1$, assuming that $\tilde \chi^0_1$  is the NLSP and Gr the LSP.
The Mathematica result is given as {\tt \{gam2tot, gam3tot\}}.

\noindent
\comarray{call stau1toGrwidth2body(gamma\_Grtau)}{Stau1toGrTauWidth[]}
calculates the decay width of $\tilde \tau_1$ decaying into ${\tilde G}\, \tau$, named {\tt gamma\_Grtau}.
Assuming that $\tilde \tau_1$ is the NLSP and ${\tilde G}$ the LSP, 
this is the only possible two-body decay of $\tilde \tau_1$.

\noindent
\comarray{Stau1toGrWidth3body(idecnumber)}{Stau1toGrWidth3body[idecnumber]}
returns a certain $\tilde \tau_1$ three-body decay to $\tilde G$ and a pair of non-SUSY particles,
 assuming that $\tilde \tau_1$ is the NLSP and ${\tilde G}$ the LSP. 
{\tt decnumber} stands for 
{\tt  \{\{Stau2GrZTau, 1\}, \{Stau2GrWmNutau, 2\}, \{Stau2Grh0Tau, 3\}, 
 \{Stau2GrHHTau, 4\}, \{Stau2GrA0Tau, 5\}, \{Stau2GrHmNutau, 6\}\} }.

\noindent
\comarray{call stau1toGrtotalwidth(gam2tot, gam3tot)}{Stau1toGrTotalWidth[]} calculates 
the total two-body decay width and the non-resonant three-body decay width into SM particle pairs
of $\tilde \tau_1$, assuming that $\tilde \tau_1$ is the NLSP and Gr the LSP. Note, that in this case
all possible $\tilde \tau_1$ three-body decay widths are non-resonant.
The Mathematica result is given as {\tt \{gam2tot, gam3tot\}}.

\noindent
\comarray{call stop1toGrwidth2body(gamma\_Grtop)}{Stop1toGrTopWidth[]}
calculates the decay width of $\tilde t_1$ decaying into ${\tilde G}\, t$, named {\tt gamma\_Grtop}.
Assuming that $\tilde t_1$ is the NLSP and ${\tilde G}$ the LSP, 
this is the only possible two-body decay of $\tilde t_1$.

\noindent
\comarray{Stop1toGrWidth3body(idecnumber)}{Stop1toGrWidth3body[idecnumber]}
returns a certain $\tilde t_1$ three-body decay to $\tilde G$ and a pair of non-SUSY particles,
 assuming that $\tilde t_1$ is the NLSP and ${\tilde G}$ the LSP. 
{\tt decnumber} stands for 
{\tt  \{\{Stop2GrZTop, 1\}, \{Stop2GrWpBottom, 2\}, \{Stop2Grh0Top, 3\}, 
 \{Stop2GrHHTop, 4\}, \{Stop2GrA0Top, 5\}, \{Stop2GrHpBottom, 6\}\} }.

\noindent
\comarray{call stop1toGrtotalwidth(gam2tot, gam3tot)}{Stop1toGrTotalWidth[]} calculates 
the total two-body decay width and the non-resonant three-body decay width into SM particle pairs
of $\tilde t_1$, assuming that $\tilde t_1$ is the NLSP and Gr the LSP. Note, that in this case
phenomenologically only the top propagator in the channels with $W^+$ or $H^+$ can become resonant. 
The Mathematica result is given as {\tt \{gam2tot, gam3tot\}}.


\section{Numerical  results}
The unstable gravitino case, using {\tt GravitinoPack}, has been discussed in \cite{Gravitino_decays}.
On the other hand, if 
the gravitino $\widetilde G$ is the stable LSP it can play the role of the DM particle.
This scenario is called gravitino DM model (GDM).  In this case other sparticles that play the role of the 
NLSP as the neutralino, stau and stop can decay to $\widetilde G$ and
Standard Model particles. The details have been already discussed in the introduction.
As basis for the numerical analysis we will use the CMSSM and pMSSM supersymmetric models,
assuming that the gravitino is stable. 

Usually in the GDM based on CMSSM (GDM/CMSSM) models the NLSP is either the 
lightest  neutralino or the stau. There is also a narrow part of the 
parameter space, especially for large value of the trilinear couplings $A_0$
favoured by the Higgs mass, where the NLSP can be the lightest stop.  These cases are 
provided as options in the {\tt GravitionoPack 1.0}. In the general pMSSM there are many more 
possibilities for NLSP. In the present analysis we use stau and stop as representative examples 
for a slepton and squark,  in particular including non-trivial mixing effects in the mass eigenstates. 
 
We start discussing the neutralino NLSP case in the context of the GDM/CMSSM.  
For this particular case we have chosen 
a benchmark point with the CMSSM parameters $m_0=1600$, $M_{1/2}=5000$, $A_0=-4000$ GeV, and
$\tanb=10$. The mass of $\tilde {\chi}^0_1$ is 2282 GeV.
The benchmark points we study in this section are compatible with the cosmological|~\cite{WMAP9,planck,xenon100} 
and LHC constraints (Higgs mass $\simeq 126 \GeV$,
 LHCb bounds for rare decays etc.)~\cite{spheno,FH}. It is worth mentioning that the gravitino DM relic density and the NLSP relic density 
 are related by
  \beq
 \frac{\Omega_{\mathrm {NLSP}}}{\Omega_{\widetilde G}}= \frac{m_{\mathrm {NLSP}} }{m_{\widetilde G}}  > 1\, .
 \eeq
The cosmological bound for the gravitino relic density can be understood as upper bound $\Omega_{\widetilde G} h^2 \leq 0.12$.
Therefore, one can have in addition gravitino production during reheating after inflation, if the reheating temperature is relatively large, 
of the order of $\sim 10^{10} \GeV$.

\begin{figure}[t!]
\mbox{\hspace{2mm}
\includegraphics[width=7.7cm]{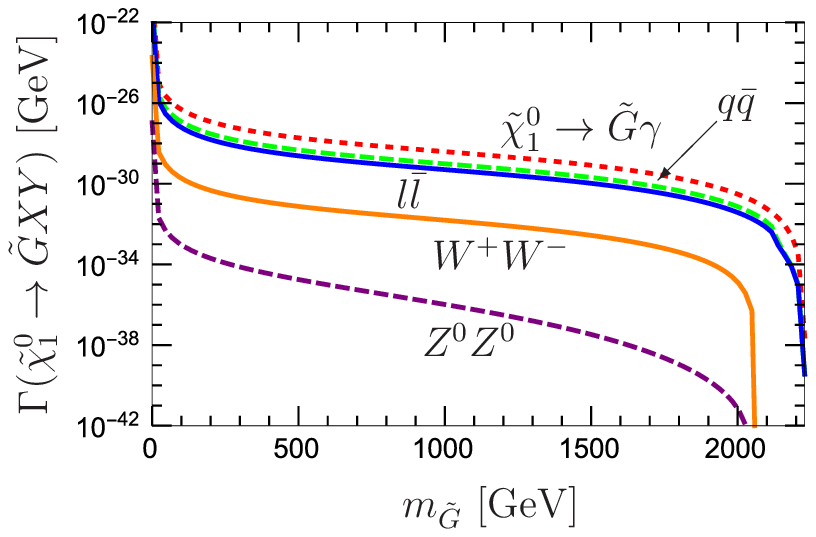}
\hspace{3mm}
\hfill
\includegraphics[width=7.6cm]{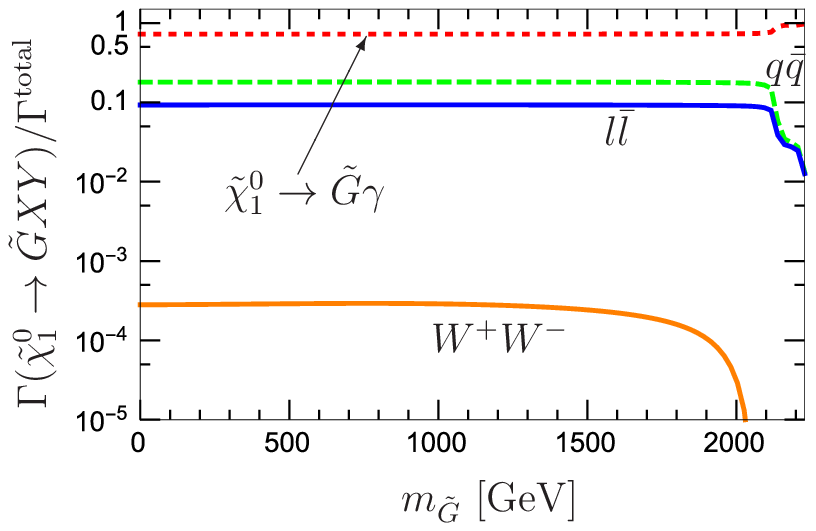}
}
\vspace*{-0.5cm}
\caption[]{The three-body decay widths of the neutralino NLSP decaying into the gravitino $\widetilde G$   and other particles, in the GDM/CMSSM scenario.  
The dominant channels  $q \bar q$, $l \bar l$, $W^+W^-$, and $ZZ$ are marked in the figure;
$q \bar q$ stands for the sum over all six quark flavors and $l  \bar l$ for the sum over the three charged lepton and three neutrino flavors.
The red dotted lines denote the two-body decay $ \lsp \to \widetilde G  \gamma$. 
 In the right figure we display the corresponding branching ratios for  the decay channels plotted in the left figure.}
\label{fig:neu_dec}
\end{figure}

\begin{figure}[h!]
\mbox{\hspace{2mm}
\includegraphics[width=7.7cm]{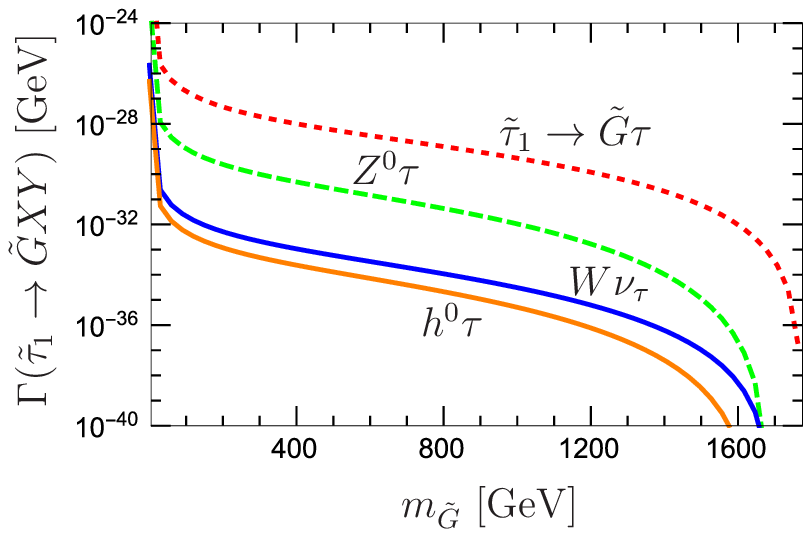}
\hspace{3mm}
\hfill
\includegraphics[width=7.6cm]{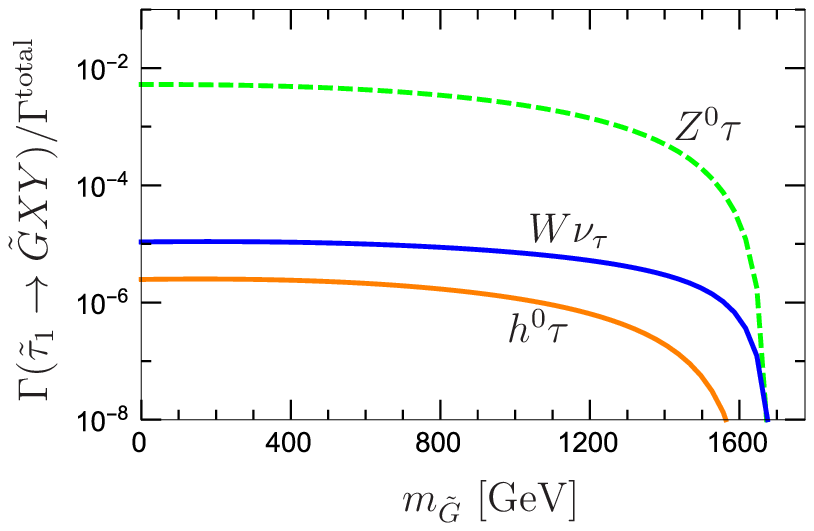}
}
\vspace*{-0.5cm}
\caption[]{The three-body decay widths of the stau  NLSP
 decaying into the gravitino $\widetilde G$  and other SM particles, in the GDM/CMSSM scenario. 
 We present the dominant two body channel $\widetilde G  \tau $ and the three-body channels 
 $ \widetilde G Z \tau  $, $ \widetilde G W^- \nu_\tau  $ and $ \widetilde G h  \tau  $.
 In the right figure we display the corresponding branching ratios for 
the decay channels plotted in the left figure except $\tilde \tau_1\to  \widetilde G \tau$.}
\label{fig:stau_dec}
\end{figure}

In Figure~\ref{fig:neu_dec} we present the corresponding decay widths (left figure) and the branching ratios (right figure) for the 
neutralino decays into $\widetilde G$   and other particles.
The dominant channels  $q \bar q$, $l \bar l$, $W^+W^-$, and $ZZ$ are marked in the figure;
$q \bar q$ stands for the sum over all six quark flavors and $l  \bar l$ for the sum over the three charged lepton and three neutrino flavors.
The red dotted lines denote the two-body decay $ \lsp \to \widetilde G  \gamma$, that dominates the neutralino decay amplitude,
as can be seen in the left panel  that illustrates the branching ratios.
On the other hand, the  three-body decay channels  $q \bar q$ and $l \bar l$ are of the order of 10\%, while 
$W^+W^-$, and $ZZ$ channels are much smaller. For this particular CMSSM point the other decay channels are even smaller. 
This happens because the neutralino is predominantly  bino,  at this particular point of the parameter space. Later we will
present cases where the higgsino components of the neutralino NLSP will boost  other channels. 

In Figure~\ref{fig:stau_dec} we present the  decays widths (left figure) and the branching ratios (right figure) for the 
stau NLSP decays into the gravitino and other particles, in the GDM scenario. 
The CMSSM parameters are $m_0=1000$, $M_{1/2}=4200$, $A_0=-2500$ GeV, and
$\tanb=10$. The mass of $\tilde\tau_1$ is 1795 GeV. The dominant decay channel is the 
two body decay $\tilde\tau  \to \widetilde G \tau $. In addition we plot the three-body channels 
$\widetilde G Z^0 \tau  $, $\widetilde G W^- \nu_\tau$, and $\widetilde G h^0  \tau$. 
The widths of the channels involving heavier Higgs bosons in the final state, are much smaller or even zero.

\begin{figure}[t!]
\mbox{\hspace{2mm}
\includegraphics[width=7.7cm]{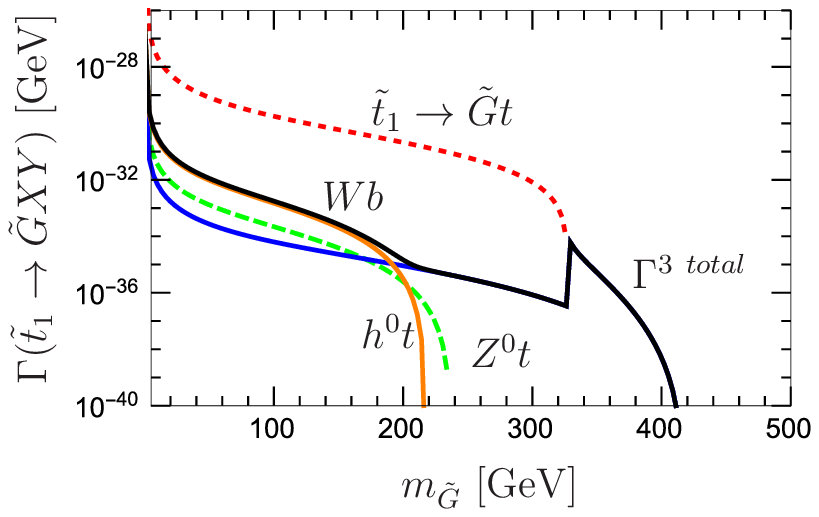}
\hspace{3mm}
\hfill
\includegraphics[width=7.6cm]{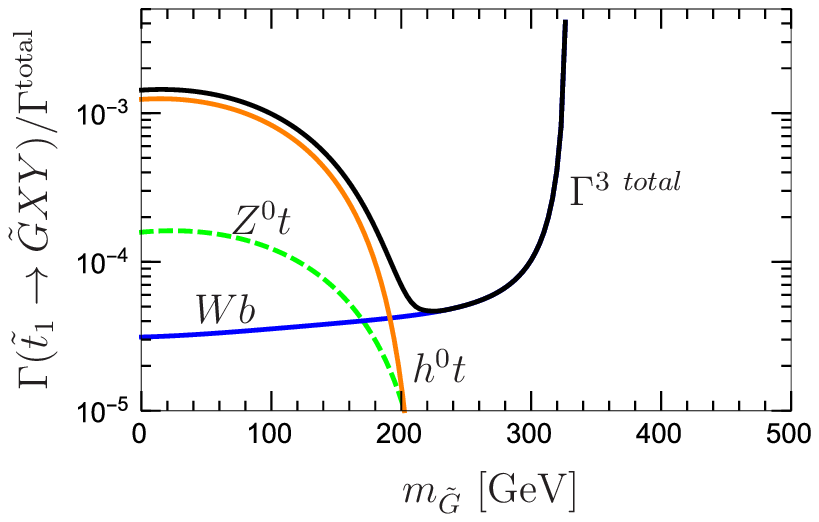}
}
\vspace*{-0.5cm}
\caption[]{The three-body decay widths of the stop  NLSP
 decaying into the gravitino $\widetilde G$  and other SM particles, in the GDM/CMSSM scenario.  
 We present the dominant two body channel $\widetilde G  \tau $ and the three-body channels 
 $ \widetilde G Z t  $, $ \widetilde G W^- b  $ and $ \widetilde G h^0  b  $.
 In the right figure we display the corresponding branching ratios for 
the decay channels plotted in the left figure except $\tilde t_1\to  \widetilde G t$.}
\label{fig:stop_dec}
\end{figure}

Similarly in Figure~\ref{fig:stop_dec} we present the  decays widths (left figure) and the branching ratios (right figure) for the 
stop NLSP decays into the gravitino $\widetilde G$ other particles. 
The CMSSM parameters are
$m_0=3000$, $M_{1/2}=1090$, $A_0=-7500$ GeV and
$\tanb=30$. The  mass of the NLSP $\tilde t_1$ is 501~GeV.
As in the case of the $\tilde \tau$ decays we present also the three-body channels 
$\widetilde G Z^0 t$, $\widetilde G W^- b$ and $\widetilde G h^0  t$. 
Again the  channels involving the heavier Higgs bosons are negligible.
The dominant decay channel is the  two body decay $\tilde t_1 \to \widetilde G t $, up to the
kinematical threshold $m_{\widetilde G} = m_{\tilde t_1} - m_t \sim 325$~GeV. As 
can been seen in both plots in Figure~\ref{fig:stop_dec}, beyond this point the 2-body channel is closed
and it dominates the 3-body channel $\widetilde G W^- b$. 
This is clearer visible in the right plot, where for $m_{\widetilde G} > m_{\tilde t} - m_t$ the $ \widetilde G W^- b$
decay channel grows after this point and eventually reach the maximum value one outside of the displayed region. 

\begin{table}[ht!]
\begin{center}
\begin{tabular}{c|c|c|c}
\hline 
  Parameters  & $\tilde \chi_1^0$ decay &   $\tilde \tau_1$ decay &  $\tilde t_1$ decay \\ 
\hline  \hline 
       $\tan\beta=  {\langle H^0_2  \rangle}/ {\langle H^0_1  \rangle}$ & 40  &  20 & 30 \\ 
       $\mu$, higgsino mixing parameter & 1  & 1.5  &  2    \\ 
       $M_A$,  $A^0$ Higgs boson mass & 2.2 & 1.5 & 2     \\ 
       ($M_1$, $M_2$, $M_3$), gauginos  masses& (1.1, 1.2, 2.8)  & (2, 3, 7) & (1, 2, 2.5)  \\ 
    $A_t$,  top  trilinear coupling &  $-$4.3  & $-$3  &  $-$4.4  \\ 
    $A_b$,  bottom  trilinear coupling & $-$6.3 & $-$3  &   $-$8 \\ 
    $A_\tau$, tau  trilinear coupling & $-$2.8  &  $-$3 &  $-$6.7    \\ 
       $m_{\widetilde{q}_{L}}$,  1$^{\rm st}$/2$^{\rm nd}$ family  $Q_L$ squark mass & 2  & 3  &  3.6 \\ 
       $m_{\widetilde{u}_{R}}$,  1$^{\rm st}$/2$^{\rm nd}$ family  $U_R$ squark & 4  &  3 &  3.6 \\  
       $m_{\widetilde{d}_{R}}$,   1$^{\rm st}$/2$^{\rm nd}$ family $D_R$ squark & 4  &  3 &  3.6  \\  
       $m_{\widetilde{\ell}_{L}}$,   1$^{\rm st}$/2$^{\rm nd}$ family  $L_L$ slepton & 2 & 2  & 3  \\ 
       $m_{\widetilde{e}_{R}}$,   1$^{\rm st}$/2$^{\rm nd}$ family  $E_R$ slepton & 4   & 2  &  3 \\  
      $m_{\widetilde{Q}_{3L}}$,  3$^{\rm rd}$family  $Q_L$ squark & 3.5 & 7  & 2.3   \\  
      $m_{\widetilde{t}_{R}}$,  3$^{\rm rd}$family  $U_R$ squark & 3.5  & 7  & 1  \\  
       $m_{\widetilde{b}_{R}}$,  3$^{\rm rd}$ family $D_R$ squark & 3.5  & 7  & 3   \\
       $m_{\widetilde{L}_{3L}}$,  3$^{\rm rd}$ family $L_L$ slepton & 1.25 & 1.2 & 2.7 \\ 
       $m_{\widetilde{\tau}_{R}}$,  3$^{\rm rd}$ family $E_R$ slepton & 3.5 & 1.2  & 2.2 \\ 
\hline         
\end{tabular}
\end{center}
\caption{The pMSSM parameters used as  input for the three scenarios in our analysis.
All values but $\tan\beta$ are given in TeV.}
\label{tab:pMSSM}
\end{table}

In addition, we will study three representative points from the pMSSM~\cite{pMSSM} supersymmetric scenario, each for the 
neutralino, stau and stop NLSP case as before. In the pMSSM model we have relaxed the unification conditions 
for the soft parameters
at the GUT scale and we used the values given in Table~\ref{tab:pMSSM}.

\begin{figure}[t!]
\mbox{\hspace{2mm}
\includegraphics[width=7.7cm]{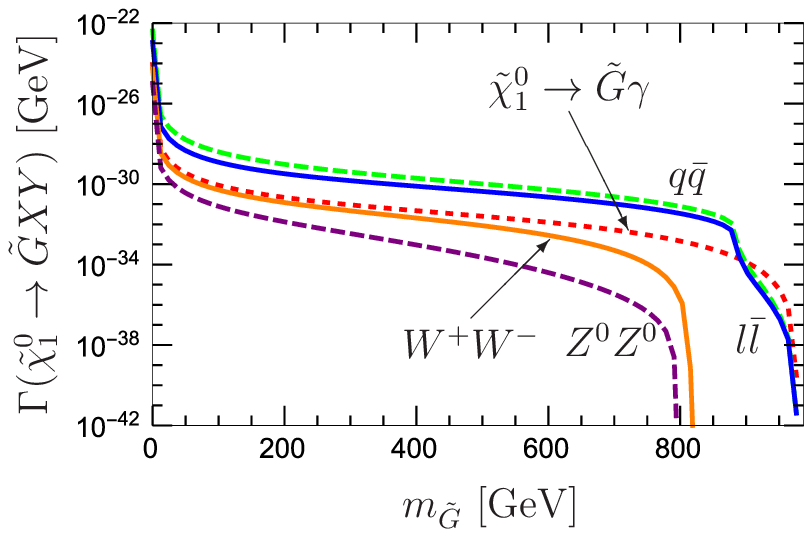}
\hspace{3mm}
\hfill
\includegraphics[width=7.6cm]{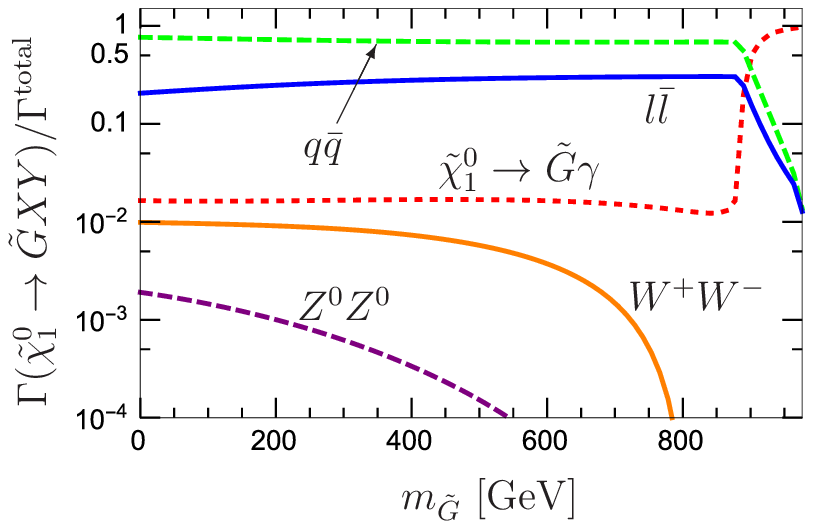}
}
\vspace*{-0.5cm}
\caption[]{The three-body decay widths of the neutralino decaying into the gravitino $\widetilde G$   and other particles, in the pMSSM scenario.  
The dominant channels  $q \bar q$, $l \bar l$, $W$-pairs, and $Z$-pairs are marked in the figure;
$q \bar q$ stands for the sum over all six quark flavors and $l  \bar l$ for the sum over the three charged lepton and three neutrino flavors.
The red dotted lines denote the two-body decay $ \lsp \to \widetilde G  \gamma$. 
In the right figure we display the corresponding branching ratios for  the decay channels plotted in the left figure.}
\label{fig:neu_dec1}
\end{figure}

In Fig.~\ref{fig:neu_dec1} we present the neutralino NLSP point, in the pMSSM. 
We have chosen  the soft SUSY parameters for this particular point
in such a way that the NLSP is predominantly higgsino, that is 
$ \lsp =   0.263\, \tilde{B} -0.210 \, \tilde{W^3}  + 0.673 \, \tilde{H_1^0} - 0.659\, \tilde{H_2^0}$.
For this reason we see that the 3-body channels $q \bar q$ and $l \bar l$ that are driven by the higgsino dominant 
$ \widetilde G \,  \lsp \, Z^0$ coupling to predominate the neutralino decay width, up to $m_{\widetilde G } \simeq 950 $ GeV where 
the 2-body channel $ \lsp \to \widetilde G  \gamma$ becomes kinematically accessible. In addition, we also present the 
$W^+ \, W^-$ and $Z^0\, Z^0$ channels that are significantly enhanced in comparison to the corresponding CMSSM scenario
presented in Figure~ \ref{fig:neu_dec}, for the same reason.

\begin{figure}[h!]
\mbox{\hspace{2mm}
\includegraphics[width=7.7cm]{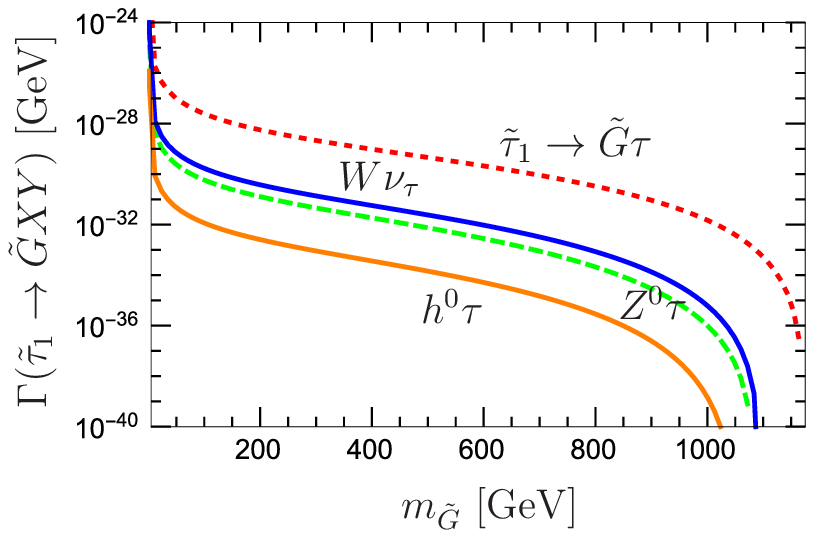}
\hspace{3mm}
\hfill
\includegraphics[width=7.6cm]{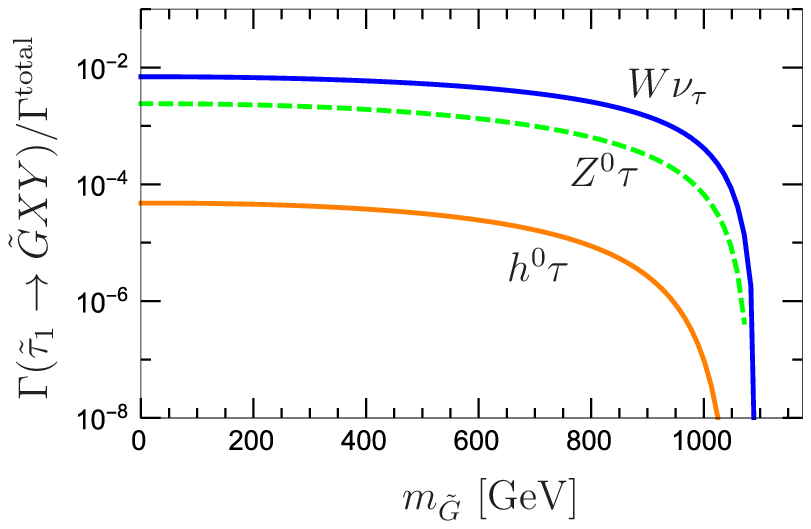}
}
\vspace*{-0.5cm}
\caption[]{The three-body decay widths of the stau NLSP
decaying into the gravitino $\widetilde G$  and other SM particles, in the pMSSM scenario. 
We present the dominant two body channel $\widetilde G  \tau $ and the three-body channels 
$\widetilde G Z^0 \tau$, $\widetilde G W^- \nu_\tau$ and $\widetilde G h^0  \tau$.
In the right figure we display the corresponding branching ratios for 
the decay channels plotted in the left figure except $\tilde \tau_1\to  \widetilde G \tau$.}
\label{fig:stau_dec1}
\end{figure}

On the other hand Figure~\ref{fig:stau_dec1} is  similar to the corresponding one of the CMSSM case presented  in 
Figure~\ref{fig:stau_dec}.  The only difference is that in the pMSSM case the $\widetilde G W^- \nu_\tau$ decay dominates,
while the $\widetilde G Z^0 \tau$ decay is by far the most dominant three-body decay in the CMSSM stau NLSP point. This is because  in the 
pMSSM point the $ \widetilde G W^- \nu_\tau $ channel is enlarged due to the sizeable $\tilde{\tau}_1\, \tilde {\chi}_1^+ \nu_\tau$ coupling.

\begin{figure}[t!]
\mbox{\hspace{2mm}
\includegraphics[width=7.7cm]{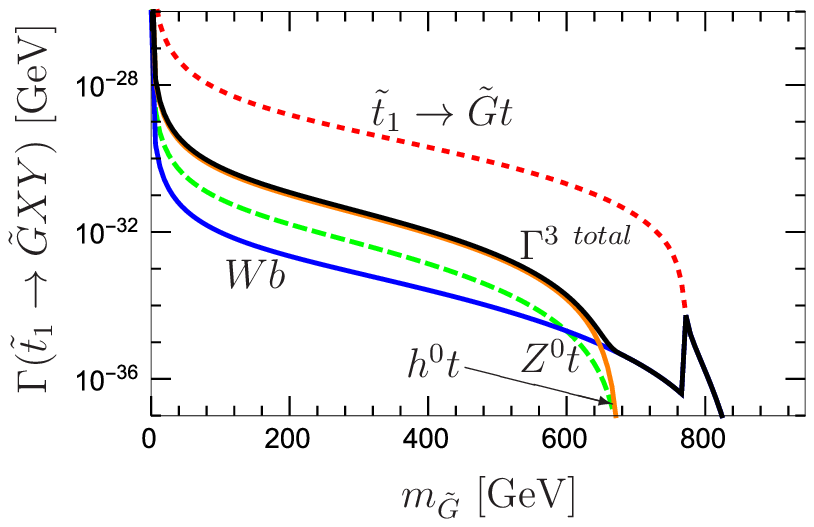}
\hspace{3mm}
\hfill
\includegraphics[width=7.6cm]{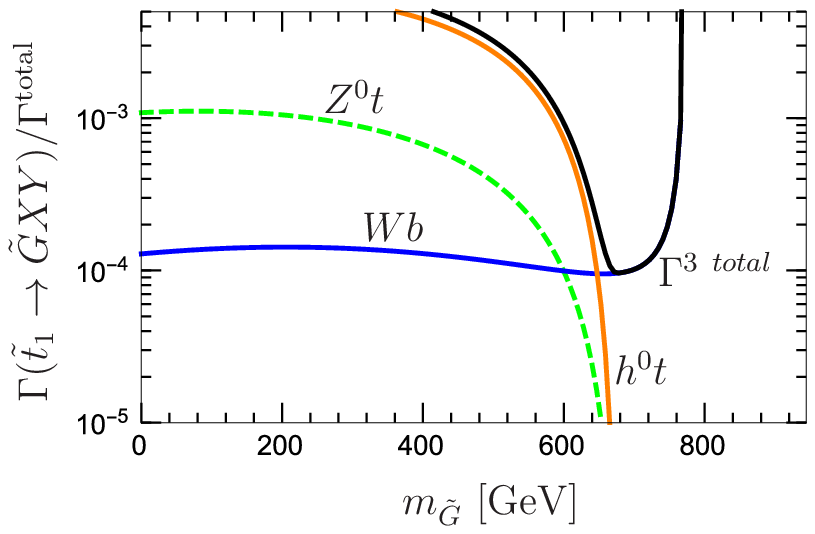}
}
\vspace*{-0.5cm}
\caption[]{The three-body decay widths of the stop NLSP
decaying into the gravitino $\widetilde G$  and other SM particles, in the pMSSM scenario.  
We present the dominant two body channel $\widetilde G  \tau $ and the three-body channels 
$ \widetilde G Z t  $, $ \widetilde G W^- b  $ and $ \widetilde G h^0  b  $.
In the right figure we display the corresponding branching ratios for 
the decay channels plotted in the left figure except $\tilde t_1\to  \widetilde G t$.}
\label{fig:stop_dec1}
\end{figure}

Finally, in Figure~\ref{fig:stop_dec1} we present the stop NLSP point in the pMSSM case. 
 For this point the $\tilde t_1$  NLSP  mass is about 830 GeV.
 As in the corresponding CMSSM case the  2-body channel $\tilde t  \to \widetilde G t $
 dominates up to the
kinematical threshold $m_{\widetilde G} = m_{\tilde t} - m_t$.
After this value of the gravitino mass the 2-body channel is closed and the 
3-body  $ \widetilde G W^- b  $ channels takes over, as has happened in the CMSSM case. 
 
To summarise the representative cases both in CMSSM and pMSSM, we can see that the full knowledge of all the two- and three-body
decay channels of the NLSP unstable particle is essential for the precise calculation of the decay width and the various 
branching ratios. This is actually the big advantage of using the {\tt GravitinoPack}, since it gives all computed results both in 
{\sc Fortran} and within the {\sc Mathematica} environment. It also supports SLHA input format.
These results enable the user of the package to apply precisely the relevant 
cosmological constraints, especially those related to the BBN predictions, to various supersymmetric models.

\section{Summary}

We have studied supersymmetric models, where the gravitino is the lightest supersymmetric particle (LSP).
In this case the gravitino can play the role of the dark matter particle. The Next to Lightest Supersymmetric Particle (NLSP)
can be either the lightest neutralino $\lsp$ or a sfermion, as stau $\tilde \tau_1$ or stop $\tilde t_1$. 
These three cases have been discussed in this work using {\tt GravitinoPack}.  
Although these cases can usually be found in the CMSSM parameter space, 
they are also representative for a slepton, squark and a gaugino NLSP in a more general supersymmetric scenario like in the pMSSM.
 
 We have calculated all two- and three-body decays of the NLSP neutralino, stau, and stop
 to the gravitino LSP and one or two SM particles. The products of these decays carry electromagnetic energy and can build hadrons
 that influence the predictions of BBN, since the gravitational nature of these decays place them in this time scale.  
 Therefore, the detailed knowledge of the relevant branching ratios and decay widths is important to study and 
 constrain various supergravity models. 
 
 To facilitate the application of BBN constraints, we have developed the public available computer tool {\tt GravitinoPack}. This
 numerical package based on an autogenerated Fortran~77 code, calculates the branching ratios and decay widths for the 
 NSLP decays, if the gravitino is stable and the DM particle. 
 On the other hand we have already presented in~\cite{Gravitino_decays} the complementary case,
 where the gravitino is unstable and can decay into a neutralino and SM particles.  Moreover, we have provided all relevant 
 technical details for its use. {\tt GravitinoPack} can be used directly at the Fortran level or more conveniently
 with {\sc Mathematica} via MathLink functions.
 
 As in the case of the decays of the unstable gravitino, the three-body decays can be important in the case of the unstable 
 neutralino, stau or stop NLSP. In Section~4 we have seen this feature especially in the region below the kinematical threshold 
 of the subleading two-body decays. Thus, {\tt GravitinoPack} provides important results
 on the decays of unstable NLSP's or gravitino, making the application of the BBN data more predictive.  
 
\section*{Acknowledgements}

This work is supported by the Austrian Science Fund (FWF) P~26338-N27
and by the European Commission through the ``HiggsTools`` Initial Training Network PITN-GA-2012-316704.
The authors thank Benjamin Fuks for the correspondence concerning the comparison 
of the two-body decays with the package FeynRules
and they are grateful to Walter Majerotto for helpful comments on the manuscript.

\begin{appendix}

\section{Gravitino Interactions with the MSSM}

\label{Gr_interactions}

\noindent
We start with the relevant supergravity Lagrangian \cite{Pradler}.
\begin{align}
  \label{eq:gravitino interaction lagrangian}
  \mathcal{L}^{(\a) }_{\grav,\,\text{int}} &= -\frac{i}{\sqrt{2}\mplanck} \left[
  \mathcal D^{(\a) }_{\m} \phi^{*i} \overline{\grav}_\n \g^\m \g^\n \cf Li
  -\mathcal D^{(\a) }_{\m} \phi^{i} \cfb Li \g^\n \g^\m  \grav_\n
\right]
 \nonumber \\
  &-  \frac{i}{8\mplanck}\overline\grav_\m [\g^\r,\g^\s] \g^\m
  {\l}^{(\a)\, a} {F}_{\r\s}^{(\a)\, a}\,,
\end{align}
with the covariant derivative given by
\begin{equation}
   \label{eq:covariant derivative}
    \mathcal{D}^{(\a)}_\mu \phi^i =
    \partial_\mu \phi^i + i {g}_\a \gbalpha {\a}{a}_\m \gen aij^{(\a)} \phi^j\, ,
\end{equation}
and the field strength tensor ${F}_{\m\n}^{(\a)\,a}$ reads
\begin{align}
  \label{eq:field strength tensor}
  {F}_{\m\n}^{(\a)\,a} & = \partial_\m {A}_\n^{(\a)\,a} -
  \partial_\n {A}_\m^{(\a)\,a} -{g}_\a f^{(\a)\,abc} {A}_\m^{(\a)\,b} {A}_\n^{(\a)\,c}\,.
\end{align}
The index $\a$ corresponds to the three groups U(1)$_Y$, SU(2)$_I$, and SU(3)$_c$ with
$a = 1, 3, 8$ and $i = 1, 2, 3$, respectively.

In detail, we get for the three covariant derivatives
\begin{eqnarray}
    \mathcal{D}^{(1)}_\mu \phi & = &
    \partial_\mu \phi + i \frac{g_1}{2} B_\mu Y(\phi) \phi\,, \nonumber \\[2mm]
\hspace*{5mm}
    \mathcal{D}^{(2)}_\mu \twovector{\phi^1}{\phi^2} & = &
   \left( \twotwomatrix{{\partial_\mu}{0}}{{0}{\partial_\mu}} +
     i \frac{g_2}{2}\twotwomatrix{{s_W A_\mu + c_W Z_\mu}{\sqrt2 W^+_\mu}}{{\sqrt2 W^-_\mu}{- s_W A_\mu - c_W
     Z_\mu}}\right)
    \twovector{\phi^1}{\phi^2}\, ,\nonumber
    \label{DmuSU_3}\\[2mm]
    \mathcal{D}^{(3)}_\mu \phi^r & = &
    \partial_\mu \phi^r + i  g_s G_\mu^a T_{rs}^{a} \phi^s \, ,
\end{eqnarray}
with $g_s$ the strong coupling, $r,s$ are color indices, and $T_{rs}^{a} = \lambda_{rs}^{a}/2$,
$\lambda_{rs}^{a}$ are
the 8 (3x3) Gell-Mann matrices. We already substituted
$B_\mu = c_W A_\mu - s_W Z_\mu\, , W^3_\mu = s_W A_\mu + c_W Z_\mu\, ,\\
W^1_\mu = (W^+_\mu + W^-_\mu)/\sqrt2\, , W^2_\mu = i (W^+_\mu - W^-_\mu)/\sqrt2$.\\
The three field strength tensors read
\begin{eqnarray}
{F}_{\m\n}^{(1)} & = & \partial_\m {B}_\n - \partial_\n {B}_\m\, ,\\
{F}_{\m\n}^{(2)\, a} & = & \partial_\m {W}^a_\n - \partial_\n {W}^a_\m  - g_2 \e^{a b c} {W}^b_\m
{W}^c_\n\, ,\\
{F}_{\m\n}^{(3)\, a} & = & \partial_\m {G}^a_\n - \partial_\n {G}^a_\m  - g_s f^{a b c} {G}^b_\m
{G}^c_\n\, ,
\end{eqnarray}
We fix $\e^{1 2 3} = 1$, $f^{a b c}$ is the structure constant of SU(3).

We have to fill in all the gauge fields, see Table~\ref{tab:gauge-fields-mssm},
and matter fields, see Table~\ref{tab:matter-fields-mssm},
of the MSSM into (\ref{eq:gravitino interaction lagrangian}).

\begin{table}[tb]
\begin{center}
\begin{tabular}{lc@{\qquad}cr@{}l}
\toprule
Name
&  Gauge bosons $A_\mu^{(\a)\,a} $ & Gauginos $\l^{(\a)\,a}$ &
 \multicolumn{2}{c}{$\big(\suthree,\sutwo\big)_{Y}$} \\
\midrule
\vspace{-0.4cm} \\
        B-boson, bino
   & $A_\mu^{(1)\,a}  = B_\m \,\d^{a1} $&
   $\l^{(1)\,a} = \widetilde{B}\, \d^{a1}  $
 & $\quad\quad(\,\mathbf{1}\,,\mathbf{1}\, ) $ &  $\vphantom{1}_{0}$
\vspace{-0.2cm} \\
\\
        W-bosons, winos
   & $A_\mu^{(2)\,a}  = W^a_\m $&
   $\l^{(2)\,a} = \widetilde{W}^a  $
 & $\quad\quad(\,\mathbf{1}\,,\mathbf{3}\, ) $ &  $\vphantom{1}_{0}$
\vspace{-0.2cm} \\
\\
        gluon, gluino
   & $A_\mu^{(3)\,a}  = G^a_\m $&
   $\l^{(3)\,a} = \widetilde{g}^a  $
 & $\quad\quad(\,\mathbf{8}\,,\mathbf{1}\, ) $ &  $\vphantom{1}_{0}$
\vspace{0.2cm}
 \\
\bottomrule
\end{tabular}
\caption[MSSM gauge fields]{Gauge fields of the MSSM
\label{tab:gauge-fields-mssm}}

\end{center}
\end{table}

%

\begin{table}[t]
\begin{center}
\begin{tabular}{ll@{\qquad}lr@{}l}
\toprule
Name
& \quad Bosons $\phi^i$ & Fermions $\chi_L^i$ &
 \multicolumn{2}{c}{$\big(\suthree,\sutwo\big)_{Y}$} \\
\midrule
\vspace{-0.4cm} \\
\multirow{3}{*}[0.3cm]{%
  \begin{minipage}{3cm}
   \mbox{Sleptons, leptons} \\ $I = 1,2,3$
  \end{minipage}
}
   & $\widetilde{L}^I =
  \begin{pmatrix}
     \widetilde{\n}^I_L\\ \widetilde{e}^{\,-\,I}_L
  \end{pmatrix}$ &
 ${L}^I =
  \begin{pmatrix}
     {\n}^I_L\\ {e}^{-\,I}_L
  \end{pmatrix}$
 & $\quad\quad(\,\mathbf{1}\,,\mathbf{2}\, ) $ &  $\vphantom{1}_{-1}$
 \vspace{0.1cm}
 \\
 & $ \widetilde{E}^{*I} = {\widetilde{e}}^{\,-\,*\,I}_R $ &
  $ E^{c\,I} ={e}^{-\,c\,I}_R $
 & $(\,{\mathbf{1}}\,,\mathbf{1}\, ) $ &  $\vphantom{1}_{+2}$
\vspace{-0.2cm} \\
\\
\multirow{4}{*}[0.56cm]{%
  \begin{minipage}{3cm}
        Squarks, quarks \\ $I = 1,2,3$ \smallskip \\ ($\times$ 3 colors)
  \end{minipage}}
 & $\widetilde{Q}^I =
  \begin{pmatrix}
     \widetilde{u}^I_L\\ \widetilde{d}^I_L
  \end{pmatrix}$ &
 ${Q}^I =
  \begin{pmatrix}
     {u}^I_L\\ {d}^I_L
  \end{pmatrix}$
 & $(\,{\mathbf{3}}\,,\mathbf{2}\, ) $ &
 $\vphantom{1}_{+\frac{1}{3}}$
\vspace{0.1cm} \\
  & $ \widetilde{U}^{*I} = \widetilde{u}^{*I}_R$ &
  $  U^{c\,I} = {u}^{c\,I}_R $
& $(\,\overline{\mathbf{3}}\,,\mathbf{1}\, ) $ &
$\vphantom{1}_{-\frac{4}{3}}$
\vspace{0.1cm}\\
  & $ \widetilde{D}^{*I} = \widetilde{d}^{*I}_R$ &
  $  D^{c\,I} = {d}^{c\,I}_R $
& $(\,\overline{\mathbf{3}}\,,\mathbf{1}\, ) $ &
$\vphantom{1}_{+\frac{2}{3}}$
\vspace{-0.2cm} \\
\\
\multirow{3}{*}[0.56cm]{%
  \begin{minipage}{3cm}
        Higgs, higgsinos
  \end{minipage}}
 &
 ${H}_d =
  \begin{pmatrix}
     H_d^0 \\ H_d^-
   \end{pmatrix}$ &
 $\widetilde{H}_d =
  \begin{pmatrix}
     \widetilde{H}^0_d\\ \widetilde{H}^-_d
  \end{pmatrix}$
& $(\,{\mathbf{1}}\,,\mathbf{2}\, ) $ &  $\vphantom{1}_{-1}$
\vspace{0.2cm}
\\
 &
 ${H}_u =
  \begin{pmatrix}
     H_u^+ \\ H_u^0
   \end{pmatrix}$ &
 $\widetilde{H}_u =
  \begin{pmatrix}
     \widetilde{H}^+_u\\ \widetilde{H}^0_u
  \end{pmatrix}$
& $(\,{\mathbf{1}}\,,\mathbf{2}\, ) $ &  $\vphantom{1}_{+1}$
\vspace{0.2cm}
 \\
\bottomrule
\end{tabular}
\caption[MSSM matter fields]{Matter fields of the MSSM
\label{tab:matter-fields-mssm}}
\end{center}
\end{table}

We start with the gravitino interaction with the Higgs bosons and higgsinos together with the SU(2) and U(1) gauginos.
First of all, the Lagrangian (\ref{eq:gravitino interaction lagrangian}) is hermitian.
This means, $(i  \mathcal  D^{(\a) }_{\m} \phi^{i} \cfb Li \g^\n \g^\m  \grav_\n)^\dagger = 
- i  \mathcal D^{(\a) }_{\m} \phi^{*i} \overline{\grav}_\n \g^\m \g^\n \cf Li$, and
the second line in (\ref{eq:gravitino interaction lagrangian}) is a real quantity.
This gives $\mathcal D^{(\a) }_{\m} \phi^{*i} = (\mathcal D^{(\a) }_{\m} \phi^{i})^\dagger$,
which is true because $\mathcal D^{(\a) }_{\m}$ is an operator which acts on $\phi^{i}$ e.g.
under SU(2), but it acts on $\phi^{*i}$ under $\overline{\rm SU(2)}$. So we can calculate first
the second term of (\ref{eq:gravitino interaction lagrangian}) and by hermitian conjugation we get
the first one. We need all the insertions for $\phi^i$, $\chi_L^i$, and $\l$'s.
The Higgs fields are $(H_d^0, H_d^-)$, and $(H_u^+, H_u^0)$ with the transformations to physical
states,
\begin{eqnarray}
H_1^1 \equiv H_d^0 & = & \frac{1}{\sqrt2}\left(v_1 + c_\a H^0 - s_\a h^0 + i ( - c_\b G^0 + s_\b A^0 \right)\, ,\\
H_1^2 \equiv H_d^- & = & - c_\b G^- + s_\b H^-\, ,\\
H_2^1 \equiv H_u^+ & = & s_\b G^+ + c_\b H^+\, ,\\
H_2^2 \equiv H_u^0 & = & \frac{1}{\sqrt2}\left(v_2 +  s_\a H^0 + c_\a h^0 + i ( s_\b G^0 + c_\b A^0
\right)\,,
\end{eqnarray}
$v_1 = v c_\b$ and $v_2= v s_\b$ using the SM convention $v  = 2 m_W /g_2$.
The Higgs superpartners, called higgsinos are left-handed by definition and transform to charginos and neutralinos as
\begin{eqnarray}
\label{tildeHtrans}
\tilde H_d^0 & = & Z^*_{k,3} P_L \tilde\chi^0_k\,, \nonumber \\
\tilde H_d^- & = & U^*_{j,2} P_L \tilde\chi^-_j\, ,\nonumber \\
\tilde H_u^+ & = & V^*_{j,2} P_L \tilde\chi^+_j\,, \nonumber \\
\tilde H_u^0 & = & Z^*_{k,4} P_L \tilde\chi^0_k\, ,
\end{eqnarray}
and the U(1) and SU(2) gauginos, which have left and right spin components, follow
\begin{eqnarray}
 \lambda^B & = & Z^*_{k,1} P_L \tilde\chi^0_k + Z_{k,1} P_R \tilde\chi^0_k\, ,\nonumber \\
 \lambda^+ & = & V^*_{j,1} P_L \tilde\chi^+_j + U_{j,1} P_R \tilde\chi^+_j\, ,\nonumber \\
 \lambda^- & = & U^*_{j,1} P_L \tilde\chi^-_j + V_{j,1} P_R \tilde\chi^-_j\, ,\nonumber \\
 \lambda^3 & = & Z^*_{k,2} P_L \tilde\chi^0_k + Z_{k,2} P_R \tilde\chi^0_k\, .
\label{lambda3transf}
\end{eqnarray}
For the second term of (\ref{eq:gravitino interaction lagrangian}) we need
$\bar\chi_L$. For next steps we need the formulas
\begin{eqnarray}
 \g^0 \g^{\mu\dagger} \g^0 = \g^\mu \quad &,& \quad
\g^0 P_{L,R}^\dagger \g^0 = P_{R,L}\, , \label{identities1} \\
C \g^{\mu\,T} C^{-1} = - \g^\mu \quad &,& \quad
C P_{L,R}^T C^{-1} = P_{L,R}\, ,\label{identities2}
\end{eqnarray}
with $\g^0 \g^0 = 1$, and for the charge conjugate matrix it holds $C^{-1} = - C = C^T$.
Applying (\ref{identities1}) we derive
$\chi_L = P_L \chi$, $\overline{\chi_L} = (P_L \chi)^\dagger \g^0 = \chi^\dagger P_L^\dagger \g^0 =
\chi^\dagger \g^0 \g^0 P_L^\dagger \g^0 = \bar \chi P_R$. We get from (\ref{tildeHtrans}) and
(\ref{lambda3transf})
\begin{eqnarray}
\label{bartildeHtrans}
\overline{\tilde H_d^0} & = & Z_{k,3} \overline{\tilde\chi^0_k} P_R\,, \quad
\overline{\lambda^B}  =  Z_{k,1}\overline{\tilde\chi^0_k} P_R  + Z^*_{k,1} \overline{\tilde\chi^0_k} P_L\, ,\  \nonumber \\
\overline{\tilde H_d^-} & = & U_{j,2} \overline{\tilde\chi^-_j} P_R\,,  \quad
\overline{\lambda^-}  =  U_{j,1} \overline{\tilde\chi^-_j} P_R + V^*_{j,1} \overline{\tilde\chi^-_j} P_L\, ,\ \nonumber \\
\overline{\tilde H_u^+} & = & V_{j,2} \overline{\tilde\chi^+_j} P_R\,,  \quad
\overline{\lambda^+}  =  V_{j,1} \overline{\tilde\chi^+_j }P_R + U^*_{j,1} \overline{\tilde\chi^+_j} P_L\, ,\ \nonumber \\
\overline{\tilde H_u^0} & = & Z_{k,4} \overline{\tilde\chi^0_k} P_R\,,  \quad
\overline{\lambda^3}  =  Z_{k,2}\overline{\tilde\chi^0_k} P_R  + Z^*_{k,2} \overline{\tilde\chi^0_k} P_L\, .
\end{eqnarray}
For the second term for the doublet $(H_d^0, H_d^-)$ we have
\begin{eqnarray}
\label{lag2H_d}
&  {\cal L}_2 \sim & {1 \over \sqrt2 \mplanck}\bigg[\left(i \partial_\mu H_d^0 - {1 \over 2}
\left( g_1 Y(H_d) B_\mu  + g_2  W^3_\mu \right)  H_d^0
- {1 \over \sqrt2} g_2 W_\mu^+ H_d^- \right) \overline{\tilde H_d^0} \g^\nu \g^\mu \grav_\n
+\nonumber\\
&& \hspace{1.2cm} \left(i \partial_\mu H_d^-  - {1   \over 2}\left( g_1 Y(H_d) B_\mu - g_2 W^3_\mu\right)  H_d^-
- {1 \over \sqrt2} g_2 W_\mu^- H_d^0 \right) \overline{\tilde H_d^-} \g^\nu \g^\mu \grav_\n \bigg]\, .
\hspace*{1cm}
\end{eqnarray}
As already mentioned, the first term we get by hermitian conjugation,
\begin{eqnarray}
\label{lag1H_d}
&   {\cal L}_1 \sim & {1 \over \sqrt2 \mplanck}\bigg[\left(-i \partial_\mu H_d^{0 *} - {1 \over 2}\left( g_1 Y(H_d) B_\mu + g_2 W^3_\mu\right) H_d^{0*}
- {1 \over \sqrt2} g_2 W_\mu^- H_d^+ \right) \overline{\grav}_\n \g^\mu \g^\nu \tilde H_d^0
+\nonumber\\
&& \hspace{1.2cm} \left(-i \partial_\mu H_d^+  - {1   \over 2}\left( g_1 Y(H_d) B_\mu - g_2 W^3_\mu \right)  H_d^+
- {1 \over \sqrt2} g_2 W_\mu^+ H_d^{0 *} \right) \overline{\grav}_\n \g^\mu \g^\nu \tilde H_d^- \bigg]\, .
\hspace*{1cm}
\end{eqnarray}
The terms for the doublet $(H_u^+, H_u^0)$ we simply get by
$H_d^0 \to H_u^+$, $H_d^- \to H_u^0$ and $Y(H_d) \to Y(H_u)$.\\
The third term is a product of two antisymmetric tensors in two Lorentz indices. Thus,
we can simplify them,
\begin{eqnarray}
\label{forlag3ew0}
\big[ \g^\r, \g^\s\big] F_{\r \s}^{(1)} & = & 2 \big[ \g^\r, \g^\s\big] \partial_\r B_\s\,,\\
\label{forlag3ew1}
\big[ \g^\r, \g^\s\big] F_{\r \s}^{(2) 1} & = & 2 \big[ \g^\r, \g^\s\big] \left(
                                 \partial_\r W^1_\s - g_2 W^2_\r W^3_\s\right)\, ,  \nonumber \\
\label{forlag3ew2}
\big[ \g^\r, \g^\s\big] F_{\r \s}^{(2) 2} & = & 2 \big[ \g^\r, \g^\s\big] \left(
                                 \partial_\r W^2_\s + g_2 W^1_\r W^3_\s\right)\, , \nonumber \\
\label{forlag3ew3}
\big[ \g^\r, \g^\s\big] F_{\r \s}^{(2) 3} & = & 2 \big[ \g^\r, \g^\s\big] \left(
                                 \partial_\r W^3_\s - g_2 W^1_\r W^2_\s\right)\, .
\end{eqnarray}
The gaugino superpartner transform analogously to the vector fields.
We need
$\l^1 = (\l^+ + \l^-)/\sqrt2\, , \l^2 = i (\l^+ - \l^-)/\sqrt2$.
Having inserted the rules for $W^{1,2}$ and $\l^{1,2}$, the third term reads
\begin{eqnarray}
& {\cal L}_3 \sim & - {i \over 4 \mplanck} \overline\grav_\m [\g^\r,\g^\s] \g^\m \Bigg[ \l^B \partial_\r B_\s
+  \l^+ \partial_\r W^-_\s + \l^- \partial_\r W^+_\s + \l^3 \partial_\r W^3_\s \nonumber\\
&&
\hspace{3.6cm} - i g_2 \l^+ W^3_\r W^-_\s + i g_2 \l^- W^3_\r W^+_\s - i g_2 \l^3 W^-_\r W^+_\s
\Bigg] \, .
\end{eqnarray}
For the Feynman rules one also needs the third term in a different form.
We know that ${F}_{\r\s}^{(\a)\, a}$ is real. Thus, it holds:
$i \overline\grav_\m [\g^\r,\g^\s] \g^\m {\l} =
(i \overline\grav_\m [\g^\r,\g^\s] \g^\m {\l})^\dagger =
-i {\l}^\dagger \g^{\m \dagger} [\g^\r,\g^\s]^\dagger  \overline\grav_\m^\dagger =
i \bar \l  \g^\m [\g^\r,\g^\s] \grav_\m$ and we get the second form
\begin{eqnarray}
& {\cal L}_3 \sim & - {i \over 4 \mplanck} \Bigg[ \partial_\r B_\s \bar\l^B
+  \partial_\r W^+_\s \bar\l^+  + \partial_\r W^-_\s \bar\l^-  + \partial_\r W^3_\s \bar\l^3 \nonumber\\
&&
\hspace{1.5cm} + i g_2 W^3_\r W^+_\s \bar\l^+ - i g_2 W^3_\r W^-_\s \bar\l^- + i g_2 W^+_\r W^-_\s \bar\l^3
\Bigg] \g^\m [\g^\r,\g^\s] \grav_\m \, . \label{lag3second}
\end{eqnarray}

Now all formulas are given to calculate the total Lagrangian without
fermion and gluino interactions.

Next we focus on the electroweak interaction with fermions and sfermions.
First of all, we need the transformations from the interaction to the physical field states, 
\begin{equation}
\tilde f_L = R^{\tilde f *}_{i1} \tilde f_i\quad , \quad
\tilde f_R = R^{\tilde f *}_{i2} \tilde f_i\, .
\label{eq:sfermionfieldtrans}
\end{equation}
Only the first line of (\ref{eq:gravitino interaction lagrangian}) is relevant.
In the following the quark or lepton flavour indices $I (= 1,2,3)$ will be suppressed.
We start with the left handed doublets $(\tilde u_L, \tilde d_L)$ and their SM partners
$(u_L, d_L)$. Recall, that $\psi_L = P_L \psi$ and thus $\overline{\psi_L} = \bar \psi P_R$.
We get from (\ref{lag2H_d}) and (\ref{lag1H_d}) by appropriate substitutions
\begin{eqnarray}
\label{lag2QL}
& {\cal L}_2 = & {1 \over \sqrt2 \mplanck}\bigg[\left(i \partial_\mu \tilde u_L - {1 \over 2}\left( g_1 Y(Q) B_\mu  + g_2 W^3_\mu \right) \tilde u_L
- {1 \over \sqrt2} g_2 W_\mu^+ \tilde d_L \right) \bar{u} \g^\nu \g^\mu P_R \grav_\n
+\nonumber\\
&& \hspace{1.2cm} \left(i \partial_\mu  \tilde d_L  - {1 \over 2}\left( g_1 Y(Q) B_\mu - g_2 W^3_\mu\right) \tilde d_L
- {1 \over \sqrt2} g_2 W_\mu^- \tilde u_L \right) \bar{d} \g^\nu \g^\mu P_R \grav_\n \bigg]\, ,
\end{eqnarray}
and
\begin{eqnarray}
\label{lag1QL}
&  {\cal L}_1 = & {1 \over \sqrt2 \mplanck}\bigg[\left(-i \partial_\mu \tilde u_L^* - {1 \over 2}\left( g_1 Y(Q) B_\mu + g_2 W^3_\mu \right) \tilde u_L^*
- {1 \over \sqrt2} g_2 W_\mu^- \tilde d_L^* \right) \overline\grav_\n \g^\mu \g^\nu P_L u
+\nonumber\\
&& \hspace{1.2cm} \left(-i \partial_\mu \tilde d_L^*   - {1 \over 2}\left( g_1 Y(Q) B_\mu - g_2 W^3_\mu \right)  \tilde d_L^*
- {1 \over \sqrt2} g_2 W_\mu^+\tilde u_L^* \right) \overline\grav_\n \g^\mu \g^\nu P_L d \bigg]\, .
\hspace{5mm}
\end{eqnarray}
For the right handed quarks the situation is more complex, because one has to work with left handed
fermion fields. We need some identities for charge conjugated spinor fields. Most important is
\begin{equation}
\psi_R^c \equiv \left(\psi_R\right)^c = \left(\psi^c\right)_L = P_L \psi^c \, .
\end{equation}
With $C P_L^T = P_L C$ the proof is
$\left(\psi_R\right)^c = C \overline{\psi_R}^T = C \left(\bar\psi P_L\right)^T = C P_L^T
\bar\psi^T = P_L C \bar\psi^T = P_L \psi^c$.
The gravitino is a Majorana particle, $\grav_\n^c = \grav_\n$. Thus, 
$\overline\grav_\n \g^\mu \g^\nu f_R^c = \bar f \g^\nu \g^\mu P_L \grav_\n$ and
$\overline{f_R^c} \g^\nu \g^\mu  \grav_\n = \overline\grav_\n \g^\mu \g^\nu P_R f$.
As $u_R$ and $d_R$ are isospin singlett fields we get
\begin{eqnarray}
\label{lag2uRdR}
& {\cal L}_2 = & {1 \over \sqrt2 \mplanck}\bigg[\left(i \partial_\mu \tilde u_R^* - {1 \over 2} g_1 Y(\tilde u^*_R) B_\mu \tilde u_R^*
\right) \overline\grav_\n \g^\mu \g^\nu P_R u
+\nonumber\\
&& \hspace{1.2cm} \left(i \partial_\mu  \tilde d_R^*  - {1 \over 2} g_1 Y(\tilde d^*_R) B_\mu  \tilde d_R^*
\right) \overline\grav_\n \g^\mu \g^\nu P_R d \bigg]\, ,
\end{eqnarray}
and
\begin{eqnarray}
\label{lag1uRdR}
&  {\cal L}_1 = & {1 \over \sqrt2 \mplanck}\bigg[\left(-i \partial_\mu \tilde u_R - {1 \over 2} g_1 Y(\tilde u^*_R) B_\mu \tilde u_R
\right) \bar u \g^\nu \g^\mu P_L \grav_\n
+\nonumber\\
&& \hspace{1.2cm} \left(-i \partial_\mu \tilde d_R - {1 \over 2} g_1 Y(\tilde d^*_R) B_\mu \tilde d_R
\right) \bar d \g^\nu \g^\mu P_L \grav_\n \bigg]\, .
\end{eqnarray}

The analogous results for the leptons we get by substitution of $u \to \nu$ and $d \to e$ in
(\ref{lag2QL}), (\ref{lag1QL}), (\ref{lag2uRdR}), and (\ref{lag1uRdR}). Note, as there is no
$\nu_R$ included, $\tilde u_R \to \tilde \nu_R \to 0$.
This means, only the second lines of (\ref{lag2uRdR}) and (\ref{lag1uRdR}) contribute in the
leptonic case.\\

Only the interactions with gluons and gluinos are still missing.
Using (\ref{DmuSU_3}) we get for the part with the left handed squarks $\tilde q_L$
\begin{eqnarray}
\label{lag2qLcolor}
& {\cal L}_2 = & {1 \over \sqrt2 \mplanck} \left(i \partial_\mu \tilde q_L^r - g_s G_\mu^a T_{rs}^{a}  \tilde q_L^s \right)
\d_{rt} \bar{q^t} \g^\nu \g^\mu P_R \grav_\n \, .
\end{eqnarray}
The matrices $T^a$ are hermitian, $T^{a *}_{rs} = T^a_{sr}$. We get
\begin{eqnarray}
\label{lag1qLcolor}
& {\cal L}_1 = & {1 \over \sqrt2 \mplanck} \left(-i \partial_\mu \tilde q_L^{r *} - g_s G_\mu^a T_{sr}^{a} \tilde q_L^{s *} \right)
\d_{rt}\overline\grav_\n \g^\mu \g^\nu P_L q^t \, .
\end{eqnarray}
For the right handed quarks we have to be careful again, because we work with the left handed
antiparticle, which transforms under $\overline{{\rm SU(3)}}$ with $- T^{a *}_{rs} = - T^a_{sr}$. We get
\begin{eqnarray}
\label{lag2qRcolor}
& {\cal L}_2 = & {1 \over \sqrt2 \mplanck} \left(i \partial_\mu \tilde q_R^{r *} + g_s G_\mu^a T_{sr}^{a} \tilde q_R^{s *} \right)
\d_{rt} \overline\grav_\n \g^\mu \g^\nu P_R q^t \, ,
\end{eqnarray}
and
\begin{eqnarray}
\label{lag1qRcolor}
& {\cal L}_1 = & {1 \over \sqrt2 \mplanck} \left(- i \partial_\mu \tilde q_R^{r} + g_s G_\mu^a T_{rs}^{a} \tilde q_R^{s} \right)
\d_{rt} \bar{q^t} \g^\nu \g^\mu P_L \grav_\n \, .
\end{eqnarray}
Similar to (\ref{forlag3ew1}-\ref{forlag3ew3}) we write
\begin{equation}
\label{forlag3qcd}
\big[ \g^\r, \g^\s\big] F_{\r \s}^{(3) a}  =   \big[ \g^\r, \g^\s\big] \left(2 \partial_\r G^a_\s -
g_s f^{a b c} G^b_\r G^c_\s \right)\, .
\end{equation}
We get
\begin{eqnarray}
\label{lag3color1}
& {\cal L}_3 = & - {i \over 8 \mplanck} \left(2 \partial_\r G^a_\s -
g_s f^{a b c} G^b_\r G^c_\s \right) \overline\grav_\m [\g^\r,\g^\s] \g^\m  \tilde g^a \, ,
\end{eqnarray}
or in a second form,
\begin{eqnarray}
\label{lag3color2}
& {\cal L}_3 = & - {i \over 8 \mplanck} \left(2 \partial_\r G^a_\s -
g_s f^{a b c} G^b_\r G^c_\s \right) \overline{\tilde g^a} \g^\m [\g^\r,\g^\s] \grav_\m \, .
\end{eqnarray}

\begin{figure}[h!]
\begin{center}
\begin{tabular}{rl}
\resizebox{7cm}{!}{\includegraphics{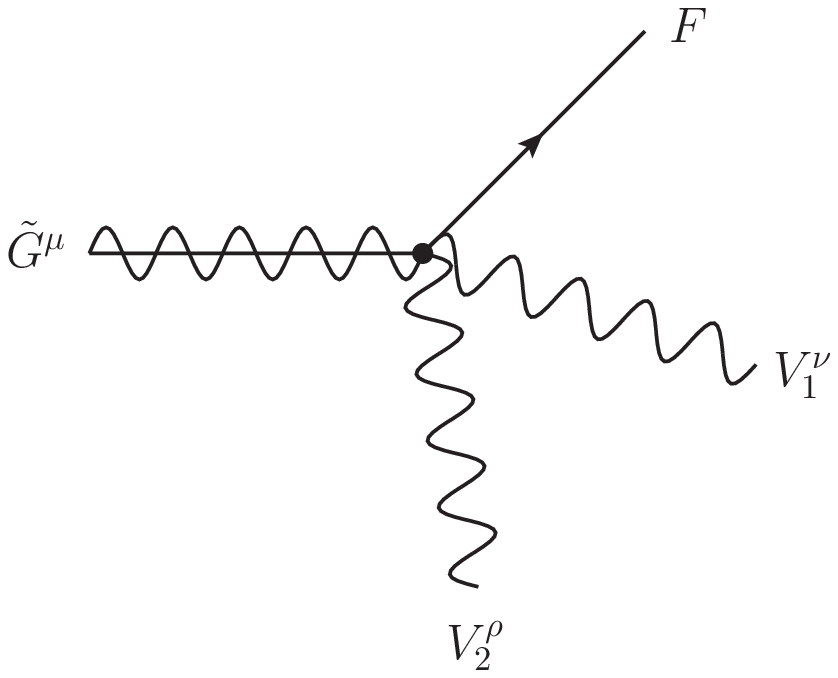}}
&
\raisebox{3cm}{
${\rm struc1:}  \fourvector{\g^\mu \left[\g^\nu,\, \g^\rho \right] P_L}{\g^\mu \left[\g^\nu,\, \g^\rho \right] P_R}
                           {\left[\g^\nu,\, \g^\rho \right] \g^\mu P_L}{\left[\g^\nu,\, \g^\rho \right] \g^\mu P_R}
. C[F|\bar F, \tilde G_\mu, V_{1\nu}, V_{2\rho}]
$
}\\
\resizebox{7cm}{!}{\includegraphics{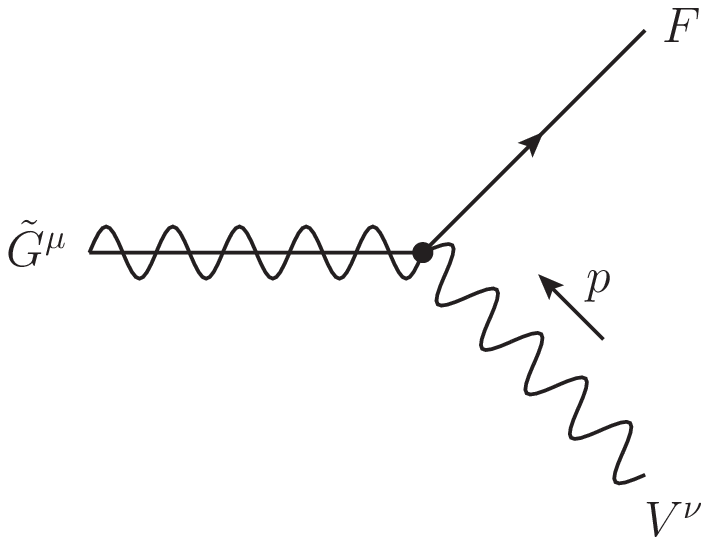}}
&
\raisebox{2.5cm}{
${\rm struc2:}  \eightvector{\g^\mu \g^\nu P_L}{\g^\mu \g^\nu P_R}
                           {\g^\mu \left[\g^\nu,\, \slash{p}\,\right] P_L}
                           {\g^\mu \left[\g^\nu,\, \slash{p}\,\right] P_R}
                           {\g^\nu \g^\mu P_L}{ \g^\nu \g^\mu P_R}
                           {\left[\g^\nu,\, \slash{p}\,\right] \g^\mu P_L}
                           {\left[ \g^\nu,\, \slash{p}\,\right] \g^\mu P_R}
. C[F|\bar F, \tilde G_\mu, V_\nu]
$}\\
\resizebox{7cm}{!}{\includegraphics{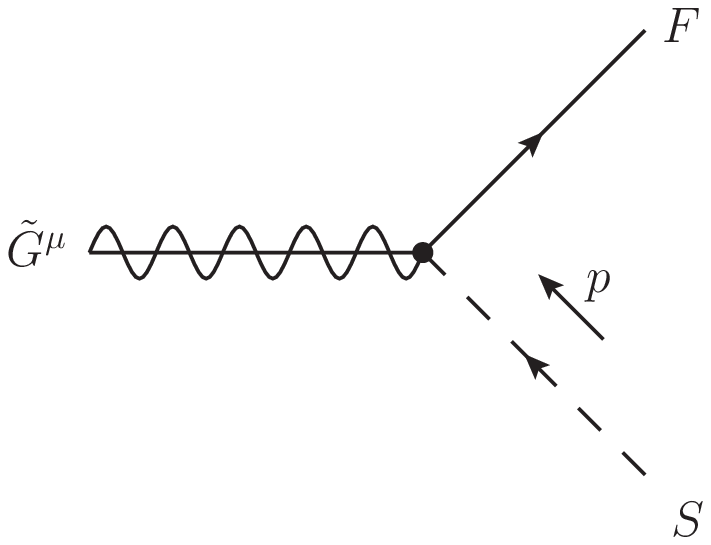}}
&
\raisebox{2.5cm}{
${\rm struc3:} \fourvector{\g^\mu \slash{p}\, P_L}{\g^\mu \slash{p}\, P_R}
                          {\slash{p} \g^\mu \, P_L}{\slash{p} \g^\mu \, P_R}
. C[F|\bar F, \tilde G_\mu, S]
$}\\
\resizebox{7cm}{!}{\includegraphics{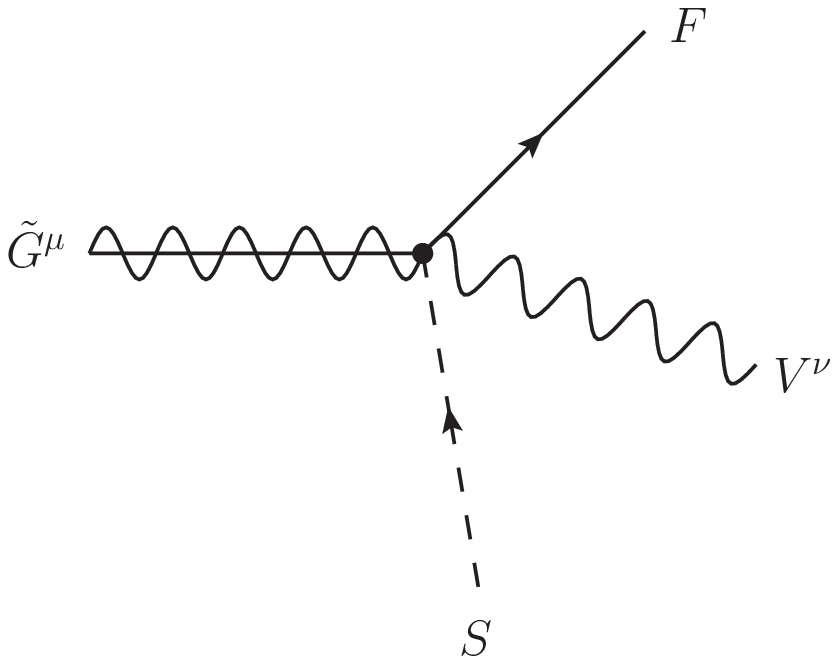}}
&
\raisebox{3cm}{
${\rm struc4:} \fourvector{\g^\mu \g^\nu P_L}{\g^\mu \g^\nu P_R}
                          {\g^\nu \g^\mu P_L}{\g^\nu \g^\mu P_R}
. C[F|\bar F, \tilde G_\mu, V_\nu, S]
$}
\end{tabular}
\end{center}
\caption{Possible structures of gravitino interactions with MSSM particles, detailed explanation
is given in the text. As all momenta are defined incoming in FA, $\partial_\mu \to -i p_\mu$. 
\label{fig:FAstructures1}}
\end{figure}

In Fig.~\ref{fig:FAstructures1} 
all possible structures are depicted for the gravitino interactions to MSSM particles in eq.~(\ref{eq:gravitino interaction lagrangian}).
We extended the FA generic Lorentz file with these structures and the 78 couplings given by the 
coupling (or coefficient) vectors $C[\ldots]$  we appended to the FA MSSM model file. 

For creating the coupling vectors of the FA model file the following relations are helpful:
\begin{eqnarray}
  \left( c \bar f \g^\m [ \g^\r \g^\s] P_{L,R} \grav_\m \right)^\dagger & = &
  - c^* \overline\grav_\m [ \g^\r \g^\s] \g^\m P_{L,R} f\,, \\
\left( c \bar f \g^\m \g^\n P_{L,R} \grav_\m  \right)^\dagger& = &
 - c^* \overline\grav_\m \g^\n \g^\m P_{R,L} f\, .
  \label{eq:helptrans1}
\end{eqnarray}
and (recall that $\grav^c_\m = \grav_\m$)
\begin{eqnarray}
   \bar{f^c} \g^\m [ \g^\r \g^\s] P_{L,R} \grav_\m & = &
   \overline\grav_\m [ \g^\r \g^\s] \g^\m  P_{R,L} f \,, \\
 \bar{f^c} \g^\m \g^\n P_{L,R} \grav_\m  & = &
 \bar \grav_\m \g^\n \g^\m P_{L,R} f\, .
\label{eq:helptrans2}
\end{eqnarray}

\begin{figure}[h!]
\begin{center}
\begin{tabular}{rl}
\resizebox{7cm}{!}{\includegraphics{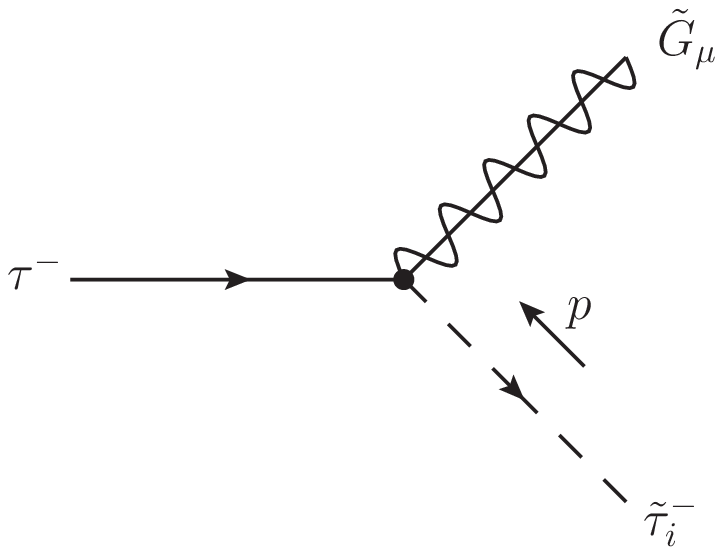}}
&
\raisebox{2.4cm}{\hspace*{-0.5cm}
$ {i \over \sqrt2 \mplanck}\, \slash{p} \g^\mu
  \left( R^{\tilde \tau}_{i2} P_R - R^{\tilde \tau}_{i1} P_L\right) $}
\end{tabular}
\end{center}
\vspace*{-0.5cm}
\caption{Feynman rule for $\tau^- \to \grav\, \tilde \tau_i^-$ derived from the Lagrangian in Eq.~(\ref{vertexGrtaustau}).
\label{fig:vertexGrtaustau}}
\end{figure}

As an illustrative example we showed in \cite{Gravitino_decays} how to get the explicit $\tilde \chi^0 \grav_\m  W^+ W^-$ interaction Lagrangian.
Here we show that for the $\grav \tilde \tau_i \tau$ interaction. From eqs.~(\ref{lag2QL}) and (\ref{lag1uRdR}) we get
\begin{equation}
 {\cal L} =  {i \over \sqrt2 \mplanck}\bigg[ (\partial_\mu \tilde\tau_L) \, \bar{\tau} \g^\nu \g^\mu P_R \grav_\n 
 - (\partial_\mu \tilde\tau_R)\,  \bar{\tau} \g^\nu \g^\mu P_L \grav_\n \bigg] + {\rm h. c.}\, .
 \end{equation}
 Writing the fields $\tilde \tau_{L,R}$ in terms of physical states by using eq.~(\ref{eq:sfermionfieldtrans}) and further using eq.~(\ref{eq:helptrans1})
 gives
\begin{equation}
 {\cal L} =  {i \over \sqrt2 \mplanck} \bigg[ (\partial_\n \tilde \tau_i)  \,\bar{\tau} \g^\m \g^\n \left( R^{\tilde \tau\,*}_{i1} P_R - R^{\tilde \tau\,*}_{i2} P_L\right) \grav_\m 
+   (\partial_\n \tilde \tau^*_i)  \, \bar \grav_\m \g^\n \g^\m  \left( R^{\tilde \tau}_{i2} P_R - R^{\tilde \tau}_{i1} P_L\right)  \tau \bigg] \, .
 \label{vertexGrtaustau}
 \end{equation} 
 Comparing this result with struc3 defined in Fig.~\ref{fig:FAstructures1} we get the coupling vectors
 \begin{equation}
 C[\bar\tau,  \tilde G^\mu, \tau_i] =  {i \over \sqrt2 \mplanck}  \fourvector{- R^{\tilde \tau\,*}_{i2}}{ \hphantom{+} R^{\tilde \tau\,*}_{i1}}{0}{0}\quad {\rm and} \quad
 C[\tau,  \tilde G^\mu, \tau^*_i] =  {i \over \sqrt2 \mplanck}  \fourvector{0}{ 0}{ - R^{\tilde \tau}_{i1}}{\hphantom{+} R^{\tilde \tau}_{i2}} \, .
 \end{equation}
 The Feynman rule for the decay $\tau^- \to \grav\, \tilde \tau_i^-$ is shown in Fig.~\ref{fig:vertexGrtaustau}.

\end{appendix}

\end{document}